\documentclass[aps,prl,twocolumn,superscriptaddress,longbibliography]{revtex4-2}

\usepackage{amsmath,amssymb,mathtools,bm}
\usepackage{graphicx}
\usepackage{xcolor}
\usepackage{tikz}
\usetikzlibrary{arrows.meta,positioning,calc,decorations.pathreplacing,fit,shapes.geometric}
\usepackage{hyperref}
\hypersetup{colorlinks=true, linkcolor=blue!60!black, citecolor=blue!60!black, urlcolor=blue!60!black}

\providecommand{\ket}[1]{|#1\rangle}
\providecommand{\bra}[1]{\langle #1|}
\providecommand{\Tr}{\mathrm{Tr}}

\begin{document}
\title{Distinct finite-temperature phase diagrams of non-invertible Kennedy--Tasaki duals}

\author{Weiguang Cao}
\email{weiguang@ust.hk}
\affiliation{Department of Physics, Hong Kong University of Science and Technology, Clear Water Bay, Hong Kong, China}
\affiliation{Center for Theoretical Condensed Matter Physics, Hong Kong University of Science and Technology, Clear Water Bay, Hong Kong, China}

\author{Haruki Watanabe}
\email{hwatanabe@ust.hk}
\affiliation{Department of Physics, Hong Kong University of Science and Technology, Clear Water Bay, Hong Kong, China}
\affiliation{Center for Theoretical Condensed Matter Physics, Hong Kong University of Science and Technology, Clear Water Bay, Hong Kong, China}
\affiliation{Institute for Advanced Study, Hong Kong University of Science and Technology, Clear Water Bay, Hong Kong, China}

\date{\today}

\begin{abstract}
Do two Hamiltonians related by a non-invertible transformation necessarily share the same finite-temperature phase diagram? For a cluster-model interpolation $H(s)$ and its Kennedy--Tasaki dual $\tilde H(s)$, the map is a partial isometry and equates only their all-plus-sector partition functions. In one dimension, thermal order is forbidden on both sides, leaving only the common zero-temperature transition at $s=\tfrac12$. In three dimensions, however, the inequivalence is exact already at $s=0$: the cluster model has an analytic paramagnetic free energy, whereas its dual $\mathbb Z_2$ gauge theory has a deconfinement transition at $T_c\approx1.31$. Quantum Monte Carlo shows that the mismatch occupies a finite window of the interpolation, marked by a self-dual frozen wedge on the cluster side and a deconfined dome on the gauge side; once both close the two phase diagrams agree again, coinciding over the entire remaining interval up to the shared trivial endpoint $s=1$.
\end{abstract}

\maketitle

\textit{Introduction.---}Dualities are thermodynamically useful because an
invertible change of variables preserves the partition function and transports
phase boundaries between different descriptions. Recent work extends symmetry
beyond groups to non-invertible operators\cite{Tachikawa:2017gyf,Bhardwaj:2017xup,Chang:2018iay,Ji:2019jhk,Thorngren:2019iar,Thorngren:2021yso}, whose generators obey fusion rules
and need not possess inverses~\cite{Aasen:2016dop,Aasen:2020jwb,Kong:2020cie,
Bhardwaj:2022yxj,Apruzzi:2022rei}. On a
lattice, such operators often arise by promoting Kramers--Wannier-type
dualities~\cite{Kaidi:2021xfk,Choi:2021kmx,Roumpedakis:2022aik,Choi:2022zal,Lootens:2021tet,Lootens:2022avn} to maps on the Hilbert space~\cite{Seiberg:2023cdc,Seiberg:2024gek,Cao:2023doz,CLY,Choi:2024rjm,Gorantla:2024ocs,Okada:2024qmk}.
Their Hamiltonian intertwining can be exact even though the map has a kernel,
and the thermodynamic content of that relation is then less immediate.

The Kennedy--Tasaki (KT) transformation provides a particularly sharp test~\cite{KT}. It
is Kramers--Wannier gauging dressed by a finite-depth cluster entangler~\cite{LOZ} and
exchanges symmetry-protected topological (SPT) order with spontaneous symmetry
breaking (SSB), while fixing the trivial paramagnet~\cite{Li:2023knf,ParayilMana:2024txy,Bhardwaj:2023bbf,Seifnashri:2024dsd,Li:2024gwx}. At zero
temperature this exchange turns a nonlocal SPT diagnostic into conventional
order. Extending the same reasoning to Gibbs states is attractive, especially
in current approaches to thermal and mixed-state SPT order
~\cite{RYKB,Ma:2022pvq,deGroot:2021vdi,LBC}, but it requires equivalence of the
thermal ensembles being compared rather than an operator intertwining alone.

For a unitary relation $H'=UHU^\dagger$, the identity
$e^{-\beta H'}=Ue^{-\beta H}U^\dagger$ guarantees that equivalence. The
non-invertible KT map studied here is instead a partial isometry onto one fusion
sector: it equates all-plus projected partition functions and annihilates the
remaining charge sectors. Omitting finitely many global sectors would change
$\ln Z$ only subextensively and would not alter a bulk phase diagram. In three
dimensions, however, the KT projector also fixes an extensive set of local
one-form charges. Restoring the discarded sectors produces an extensive
sector entropy, so their thermal occupation can change the dominant free
energy and its singularities.

We realize this mechanism in the
$\mathbb Z_2^{(0)}\times\mathbb Z_2^{(0)}$ cluster chain and the
$\mathbb Z_2^{(1)}\times\mathbb Z_2^{(1)}$
Raussendorf--Bravyi--Harrington (RBH) model~\cite{RBH,RYKB}. In one dimension
both models remain disordered at every $T>0$, so their phase
diagrams happen to agree. In three dimensions the inequivalence is exact
already at $s=0$: the cluster model has an analytic paramagnetic free energy,
whereas its dual gauge theory undergoes a deconfinement transition. Quantum
Monte Carlo (QMC) shows that this mismatch persists over a finite region of the interpolation
and is not a fine-tuned endpoint effect. The comparison separates the algebraic
statement supplied by the KT map from the dynamical input needed for
thermodynamics: pointlike domain walls proliferate immediately in one
dimension, whereas three-dimensional flux loops retain a line tension and
support a finite-temperature deconfined phase.

\textit{Fixed points and the Kennedy--Tasaki map.---}We work with qubits, $X_i,Z_i$ the Pauli operators on site $i$. The SPT fixed point, the trivial paramagnet, and their symmetry-breaking dual are
\begin{equation}
H_{\rm SPT}\!\coloneqq-\textstyle\sum_i X_iB_i,
H_{\rm para}\!\coloneqq-\textstyle\sum_i X_i,
H_{\rm SSB}\!\coloneqq-\textstyle\sum_i B_i,
\end{equation}
where each $B_i$ is a finite product of neighboring $Z$ operators and a local interaction term of $H_{\rm SSB}$. A finite-depth diagonal entangler obeys $UX_iU^\dagger=X_iB_i$, hence $H_{\rm SPT}=UH_{\rm para}U^\dagger$~\footnote{Although $U$ is finite-depth and commutes with the global symmetry ($US_aU^\dagger=S_a$), it is not symmetric gate by gate: its individual controlled-$Z$ gates break the symmetry. Its anomalous action at an open boundary therefore does not provide a symmetric finite-depth trivialization of the SPT phase.}. All three Hamiltonians have the same symmetry~\footnote{$\mathbb Z_2^{(0)}\times\mathbb Z_2^{(0)}$ in (1+1)d and $\mathbb Z_2^{(1)}\times\mathbb Z_2^{(1)}$ in (3+1)d.}, generated by products $S_a$ of $X$ operators for which the dressings cancel. We compare
\begin{equation}
\begin{aligned}
H(s)&\coloneqq(1-s)H_{\rm SPT}+s\,H_{\rm para},\\
\tilde H(s)&\coloneqq(1-s)H_{\rm SSB}+s\,H_{\rm para},
\end{aligned}\qquad s\in[0,1],
\end{equation}
where $s$ interpolates between the two fixed points.
The unitary reflection $U^\dagger H(s)U=H(1-s)$ pins the direct zero-temperature SPT--trivial transition to $s=\tfrac12$~\cite{Pivot,PivotModels}. The endpoints of $\tilde H$ are not isospectral; its corresponding reflection is non-invertible and applies only within one fusion sector.

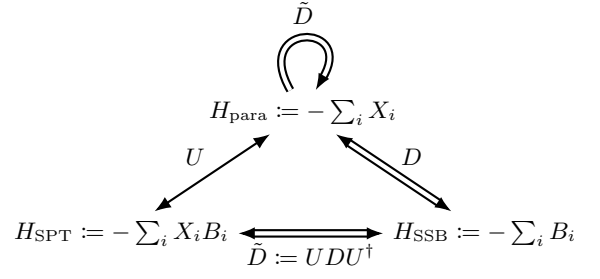
\begin{figure}[t]
\centering
\begin{tikzpicture}[>={Latex[length=2.2mm]},thick]
  \node (para) at (0,1.6)  {$H_{\rm para}\coloneqq-\sum_i X_i$};
  \node (SPT)  at (-2.4,0) {$H_{\rm SPT}\coloneqq-\sum_i X_iB_i$};
  \node (SSB)  at (2.4,0)  {$H_{\rm SSB}\coloneqq-\sum_i B_i$};
  \draw[<->] (para) -- node[above left=-1pt]{$U$} (SPT);
  \draw[<->,double,double distance=1.6pt] (para) -- node[above right=-1pt]{$D$} (SSB);
  \draw[<->,double,double distance=1.6pt] (SPT) -- node[below]{$\tilde D\coloneqq U D U^\dagger$} (SSB);
  \draw[->,double,double distance=1.6pt] (para) to[out=120,in=60,looseness=8] node[above=2pt]{$\tilde D$} (para);
\end{tikzpicture}
\caption{The unitary $U$ relates the paramagnet and SPT. The gauging $D$ and KT map $\tilde D\coloneqq UDU^\dagger$ are non-invertible partial isometries onto the all-plus sector (double lines).}
\label{fig:general}
\end{figure}

Let $D$ be the Kramers--Wannier gauging of the dressings, $DX_i=B_iD$ and $DB_i=X_iD$, with the explicit bilinear phase-map representation~\cite{Li:2024eih}
\begin{equation}
D\coloneqq\mathcal N^{-1}\!\!\sum_{\{\sigma\},\{\sigma'\}}\!(-1)^{\sum_j a_j(\sigma)\,\sigma'_j}\,\ket{\{\sigma\}}\bra{\{\sigma'\}},
\end{equation}
where $Z_i\ket{\{\sigma\}}=(-1)^{\sigma_i}\ket{\{\sigma\}}$, $\sigma_i\in\{0,1\}$, and
$a_j(\sigma)\coloneqq\sum_{k:\,Z_k\ {\rm occurs\ in}\ B_j}\sigma_k$.
Thus $B_j\ket{\{\sigma\}}=(-1)^{a_j(\sigma)}\ket{\{\sigma\}}$. The symmetric incidence matrix $M_{ij}\coloneqq1$ when $Z_j$ occurs in $B_i$ obeys $M=M^\top$, and $\mathcal N\coloneqq(2^N|\ker M|)^{1/2}$, where the total qubit number is $N=2L$ in one dimension and $N=6L^3$ in three dimensions. Dressing the gauging by the entangler gives $\tilde D\coloneqq UDU^\dagger$, with $\tilde D X_i=X_i\tilde D$ and $\tilde D X_iB_i=B_i\tilde D$ (Fig.~\ref{fig:general}), hence
\begin{equation}\label{eq:TY}
\tilde D\,H(s)=\tilde H(s)\,\tilde D,\qquad \tilde D^\dagger\tilde D=\tilde D\tilde D^\dagger=P_+.
\end{equation}
To verify the second relation, write $a(\sigma)=M\sigma$ in additive $\mathbb F_2$ notation. Contracting the intermediate basis state in $D^\dagger D$ enforces $M(\sigma+\tau)=0$, so $\bm n\coloneqq\sigma+\tau$ ranges over $\ker M$. With the stated normalization,
\begin{equation}
P_+\coloneqq\frac{1}{|\ker M|}\sum_{\bm n\in\ker M}X^{\bm n}=\prod_a\tfrac12\big(1+S_a\big)
\end{equation}
where $X^{\bm n}\coloneqq\prod_iX_i^{n_i}$ and the independent $\{S_a\}$ generate $\{X^{\bm n}\}_{\bm n\in\ker M}$, including noncontractible representatives on a torus. Symmetry of $M$ also makes $D=D^\dagger$, so both orderings give the same $P_+$. Thus the map is a partial isometry, not a unitary, onto the all-plus sector of the full fusion group. Since $US_aU^\dagger=S_a$, the projector is shared by $D$ and $\tilde D$~\cite{LOZ,KYH}; Eq.~(\ref{eq:TY}) is their Tambara--Yamagami fusion rule.

For the ground-state branches considered below, the lowest states can be
chosen in this all-plus sector (Sec.~\ref{supp:sm:setup}). Restricted to that sector,
$\tilde D$ is unitary and Eq.~(\ref{eq:TY}) maps the ground-state branches and
their sector-resolved excitation gaps across the interpolation. The nontrivial SPT phase therefore corresponds at $T=0$
to the symmetry-breaking phase, and the trivial paramagnetic phase to the
symmetry-unbroken phase, with the transition at
$s=\tfrac12$ on both sides. Non-invertibility does not obstruct this
zero-temperature phase-diagram correspondence because only the lowest state in
each phase is needed; the distinction becomes consequential when a thermal
trace populates the sectors in the kernel.

\textit{What survives at finite temperature.---}For a unitary intertwiner one can cancel the map and conclude equality of Gibbs states. For $\tilde D$ the power-series expansion of the exponential gives
\begin{equation}\label{eq:gibbs-intertwine}
\tilde D\,e^{-\beta H(s)}=e^{-\beta\tilde H(s)}\,\tilde D .
\end{equation}
Because $\tilde D$ has a nontrivial kernel, Eq.~(\ref{eq:gibbs-intertwine}) cannot be inverted. Multiplication on the right by $\tilde D^\dagger$ instead gives
\(\tilde D e^{-\beta H(s)}\tilde D^\dagger=e^{-\beta\tilde H(s)}P_+\).
Taking the trace and using cyclicity gives, for every $s$ and $T$,
\begin{equation}
Z_+(s,\beta)\coloneqq\Tr\!\big(P_+e^{-\beta H(s)}\big)=\Tr\!\big(P_+e^{-\beta\tilde H(s)}\big).
\label{eq:star}
\end{equation}
Since $P_+$ commutes with both Hamiltonians, the normalized projected states obey
\begin{equation}\label{eq:rho-intertwine}
\frac{P_+e^{-\beta\tilde H(s)}P_+}{Z_+(s,\beta)}
=\tilde D\,\frac{P_+e^{-\beta H(s)}P_+}{Z_+(s,\beta)}\,\tilde D^\dagger ,
\end{equation}
within the all-plus sector, where the restriction of $\tilde D$ is unitary. Outside its image, $\tilde D$ annihilates charge sectors rather than mapping them. Consequently Eqs.~(\ref{eq:star}) and (\ref{eq:rho-intertwine}) constrain neither the spectra outside the all-plus sector nor the Gibbs states. Thermal occupation of the annihilated sectors can therefore change the free-energy singularities.

In symmetry language, this is the weak--strong distinction emphasized in Ref.~\cite{RYKB}. For a $\mathbb Z_2$ operator $R$, weak symmetry is $R\rho R^\dagger=\rho$, while strong symmetry $R\rho=\pm\rho$ fixes one charge sector. A Gibbs state is only weakly symmetric; the KT fusion rule instead selects
\begin{equation}
\rho_+(s)\coloneqq\frac{P_+\,e^{-\beta H(s)}\,P_+}{Z_+(s,\beta)},
\qquad S_a\rho_+(s)=\rho_+(s),
\end{equation}
which is strong under the full fusion group. Strong symmetry is group-relative; in three dimensions the full group differs from the locally generated contractible group, as specified below. The exact KT identity always concerns the former.

At zero temperature the sector relation above is sufficient to match the
phases. At finite temperature, however, the full partition function
sums all charge sectors, including the exponentially many states annihilated
by $\tilde D$. Whether those states change the bulk free energy is a dynamical
question, not a consequence of the fusion algebra. The one- and
three-dimensional examples now give the two possible outcomes.

\textit{One dimension: coincident phase diagrams.---}On a ring of $L$ vertices and $L$ links, with a qubit on each,
$B_v\coloneqq\prod_{l\ni v}Z_l$ and $B_l\coloneqq\prod_{v\in\partial l}Z_v$ define the $\mathbb Z_2\times\mathbb Z_2$ cluster chain; its KT dual is two decoupled transverse-field Ising chains. The generators are
$S_{\rm vtx}^{(0)}\coloneqq\prod_vX_v$ and $S_{\rm lnk}^{(0)}\coloneqq\prod_lX_l$.
Threading an $S_{\rm vtx}^{(0)}$ flux through one bond and measuring $S_{\rm lnk}^{(0)}$ diagnoses the SPT~\cite{PT}. The exact free-fermion solution gives (Sec.~\ref{supp:sm:1d})
\begin{equation}
\begin{aligned}
\langle S_{\rm lnk}^{(0)}\rangle_{\rm tw}&=\tanh\!\big[\beta(2s-1)\big]\!\prod_{n=1}^{L-1}\!\tanh\frac{\beta\,\varepsilon_{k_n}(s)}{2},\\
\frac{\varepsilon_k(s)}{2}&\coloneqq\sqrt{(1-s)^2+s^2-2s(1-s)\cos k},
\end{aligned}
\end{equation}
where the twist selects periodic momenta $k_n\coloneqq2\pi n/L$. The isolated $k=0$ factor is signed: at $T=0$ it gives
$\langle S_{\rm lnk}^{(0)}\rangle_{\rm tw}=\mathrm{sgn}(s-\tfrac12)$ for $s\ne\tfrac12$, the nontrivial
$H^2(\mathbb Z_2\times\mathbb Z_2,U(1))=\mathbb Z_2$ response in which a vertex flux binds a link charge. At $s=\tfrac12$ the zero mode vanishes, so the canonical twisted trace is zero. For every fixed $T>0$, the logarithm of the remaining product becomes a Riemann sum, so the product's magnitude obeys the perimeter law $e^{-L/\xi}$ with
\begin{equation}\label{eq:xi1d}
\xi^{-1}\coloneqq-\frac{1}{2\pi}\int_0^{2\pi}\!dk\,\ln\tanh\frac{\beta\varepsilon_k(s)}{2}>0,
\end{equation}
finite $\xi$, and therefore vanishes as $L\to\infty$. The finite-$L$ sign still remembers the two zero-temperature phases, but it multiplies a vanishing magnitude and is not a thermodynamic invariant. Consistently, the free-fermion free energy is analytic for every $T>0$.

\begin{figure}[t]
\centering
\includegraphics[width=\columnwidth]{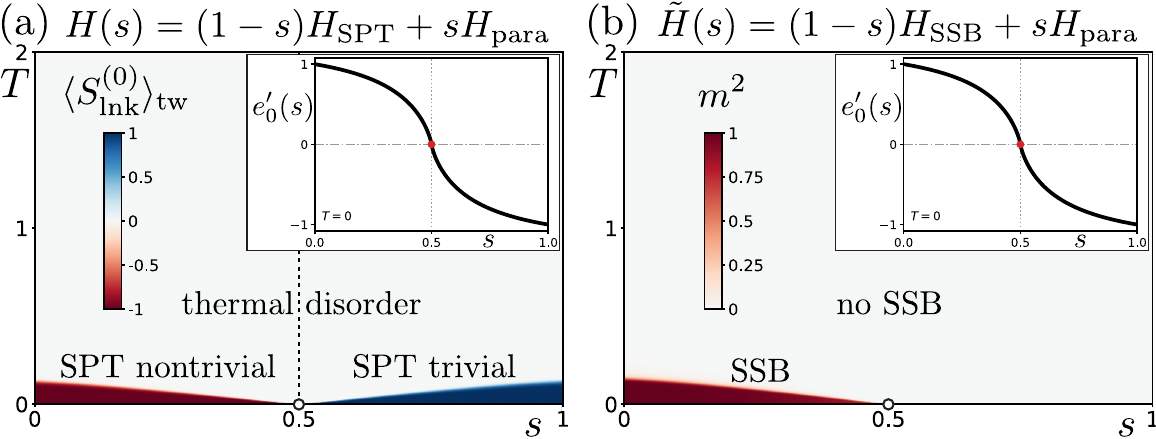}
\caption{One-dimensional finite-size crossover maps on rings of $2\times10^6$ qubits ($L=10^6$). (a)~Exact flux-twist product $\langle S_{\rm lnk}^{(0)}\rangle_{\rm tw}$ for the cluster model. (b)~Squared magnetization $m^2$ of its KT dual, from the controlled Toeplitz evaluation described in the SM (Sec.~\ref{supp:sm:1d}). Both apparent ordered regions contract onto the $T=0$ axis as $L\to\infty$. Inset: the slope $e_0'(s)$ of the ground-state energy density $e_0(s)$, which vanishes \emph{continuously} at $s=\tfrac12$ with no jump, as $e_0'(s)\sim\delta\ln(1/|\delta|)$ for $\delta\coloneqq1-2s$, while $e_0''(s)$ diverges logarithmically; see the SM (Sec.~\ref{supp:sm:1d}).}
\label{fig:1d}
\end{figure}

On the dual side the order parameter of either sublattice is
\begin{equation}
m^2\coloneqq\Big\langle\Big(\frac1L\sum_vZ_v\Big)^{\!2}\Big\rangle
=\frac1{L^2}\sum_{v,v'}\langle Z_vZ_{v'}\rangle .
\end{equation}
Jordan--Wigner transformation expresses $\langle Z_vZ_{v'}\rangle$ as a Toeplitz determinant with the same dispersion. At $T=0$ it gives
$\lim_{L\to\infty}m^2=[1-(s/(1-s))^2]^{1/4}$ for $s<\tfrac12$~\cite{Pfeuty,BarouchMcCoy}. At every $T>0$, the Toeplitz symbol has modulus below one, and the strong Szeg\H{o} theorem forces exponential decay with the same $\xi$ on the ordered side. Consequently
$m^2=L^{-1}\sum_v\langle Z_0Z_v\rangle\to0$, because the sum remains finite, as also required by the one-dimensional domain-wall argument. On the paramagnetic side the zero-temperature quantum correlation length remains finite; Eq.~(\ref{eq:xi1d}) continues to control the twist product, not that distinct quantum length.

Neither model therefore orders at finite temperature. After
matching the SPT and SSB phase labels, both transition loci consist only of
$(s,T)=(\tfrac12,0)$. The low-temperature lobes visible at finite $L$
[Fig.~\ref{fig:1d}] shrink only logarithmically with size and are crossovers,
not phases. In this example the all-plus equality extends accidentally to the
observable phase diagrams because thermal order is forbidden on both sides.
The exact Jordan--Wigner and Toeplitz derivations and the construction of
Fig.~\ref{fig:1d} are given in the SM (Sec.~\ref{supp:sm:1d}).

\textit{Three dimensions: inequivalent phase diagrams.---}On the $3$-torus of linear size $L$, put qubits on links $l$ and plaquettes $p$ and define $B_l\coloneqq\prod_{p\ni l}Z_p$ and $B_p\coloneqq\prod_{l\in\partial p}Z_l$. This gives the RBH cluster model~\cite{RBH}, protected, following Ref.~\cite{RYKB}, by the contractible one-form generators
\begin{equation}
S_c^{(1)}\coloneqq\!\!\prod_{p\in\partial c}\!\!X_p,\qquad S_v^{(1)}\coloneqq\!\!\prod_{l\ni v}\!\!X_l,
\end{equation}
whose $B$-dressings cancel pairwise. These generate only a subgroup of the symmetry: the Hamiltonian in fact commutes with the full one-form symmetry, so the six noncontractible membrane generators $S_a^{\rm nc}$ are Hamiltonian symmetries as well, and together with the contractible ones they form the KT fusion group. The contractible subgroup provides the SPT protection~\cite{RYKB}; the noncontractible generators are what the twisted-membrane index probes. Its all-plus projector, already fixed by $\tilde D^\dagger\tilde D=P_+$, is therefore
\begin{equation}
P_+=\prod_c\frac{1+S_c^{(1)}}2
\prod_v\frac{1+S_v^{(1)}}2
\prod_{a=1}^{6}\frac{1+S_a^{\rm nc}}2 .
\end{equation}
Products are over $\mathbb Z_2$: two parallel representatives of the same noncontractible plane class combine to a contractible element. Hence $\rho_+(s)$ is strong for the full fusion group.

\begin{figure}[t]
\centering
\includegraphics[width=\columnwidth]{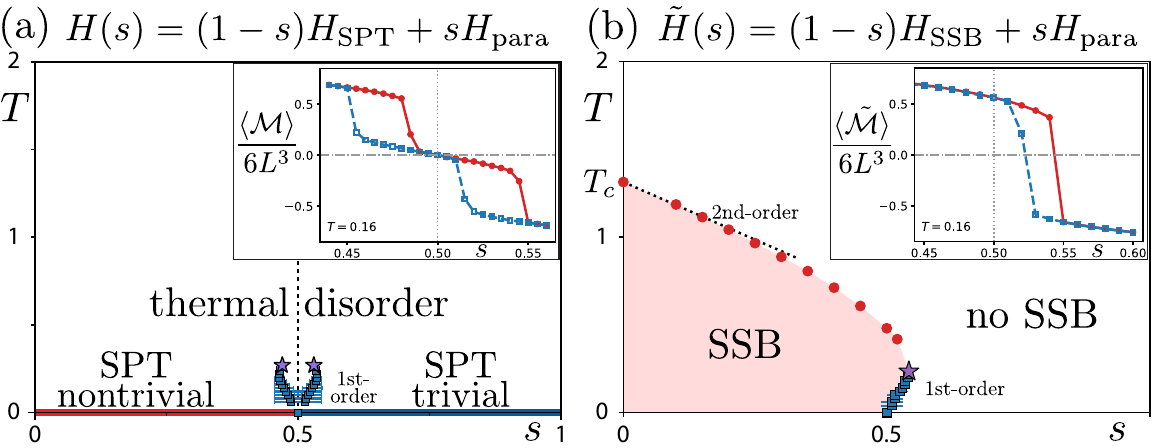}
\caption{Finite-temperature phase diagrams of (a)~the three-dimensional cluster model and (b)~its dual gauge model. In (a), the red (blue) $T=0$ segment is the exact twisted-membrane response $-1$ ($+1$); in (b), the dotted line is the exact tangent $T_c(0)(1-s)$. QMC bars denote metastability intervals, not equilibrium error bars; stars mark finite-size metastability endpoints. Insets show hysteresis at $T=0.16$. The panel-(b) density $\langle\tilde{\mathcal M}\rangle/(6L^3)$ follows from the one-copy measurement $\langle B_p\rangle-\langle X_l\rangle$ by copy symmetry.}
\label{fig:3d}
\end{figure}

The KT dual $\tilde H(s)$ is two decoupled three-dimensional $\mathbb Z_2$ gauge theories in a transverse field. Equation~(\ref{eq:star}) equates its $P_+$ trace with that of $H(s)$, but the finite-temperature phase diagrams obtained from their Gibbs states differ [Fig.~\ref{fig:3d}]. Already at $s=0$,
\(H_{\rm SPT}=UH_{\rm para}U^\dagger\), so
\(\Tr e^{-\beta H_{\rm SPT}}=(2\cosh\beta)^{6L^3}\) is analytic for every finite temperature. Its dual
\(\tilde H(0)=H_{\rm SSB}\), however, is two gauge theories whose free energy is nonanalytic at the deconfinement transition $T_c(0)\approx1.31$. This exactly solvable point alone proves that non-invertible KT duality need not preserve a finite-temperature phase diagram; the rest of Fig.~\ref{fig:3d} shows how the mismatch evolves away from it.

At this endpoint, Eq.~(\ref{eq:star}) instead concerns a thermodynamically
constrained ensemble. In three dimensions $P_+$ fixes $O(L^3)$ contractible
charges in addition to the six noncontractible ones, and the projected cluster
and gauge descriptions share the same singular free energy. At $s=0$, the
full cluster trace restores exponentially many charge sectors and
reduces, through $U$, to the analytic free-paramagnet result. Their entropy
changes $\ln Z$ at order $L^3$, so the mismatch survives the thermodynamic
limit rather than being a finite-size boundary-condition effect.
The projected phase diagram and its topological characterization are analyzed
in the companion Letter~\cite{CompanionII}.

To diagnose the SPT response directly, insert a one-form symmetry twist and measure a noncontractible membrane. On a noncontractible $2$-cycle $\Sigma$ of plaquettes ($\partial\Sigma=0$) the measured operator is
\begin{equation}
S_\Sigma^{(1)}\coloneqq\prod_{p\in\Sigma}X_p=\prod_{p\in\Sigma}X_pB_p,
\end{equation}
a product of stabilizers because $\prod_{p\in\Sigma}B_p=I$. Thread a link-symmetry flux by inserting a two-cocycle $\eta_p\in\{\pm1\}$ as $X_pB_p\to\eta_pX_pB_p$. For a noncontractible loop $\gamma$, set $\eta_p\coloneqq-1$ on the plaquettes pierced by $\gamma$ and $\eta_p\coloneqq+1$ elsewhere. Then $\prod_{p\in\partial c}\eta_p=1$ for every cube, while $[\eta]\in H^2(T^3;\mathbb Z_2)$ is the nontrivial class Poincar\'e-dual to $[\gamma]$. The resulting twisted interpolation is
\begin{equation}
\begin{split}
H_{\rm tw}(s)\coloneqq{}&-(1-s)\bigl(\textstyle\sum_l X_lB_l+\sum_p\eta_p X_pB_p\bigr)\\
&-s\bigl(\textstyle\sum_l X_l+\sum_p X_p\bigr).
\end{split}
\end{equation}
Define
\(\langle O\rangle_{\rm tw}\coloneqq\Tr[Oe^{-\beta H_{\rm tw}(s)}]/\Tr e^{-\beta H_{\rm tw}(s)}\).
The twist is absent from $H_{\rm para}$, giving $+1$ in its ground state, whereas the cluster ground state gives the braiding phase
\(\langle S_\Sigma^{(1)}\rangle_{\rm tw}=(-1)^{\Sigma\cdot\gamma}\).
Here $\Sigma\cdot\gamma$ is the mod-$2$ intersection number, so this sign is the mixed-anomaly response of the two one-form symmetries. Moreover
\([S_\Sigma^{(1)},H(s)]=0\), making the membrane eigenvalue a sector label. At $T=0$ a gapped branch stays in one such sector; a twisted reflection exchanges the endpoints and gives, for odd $\Sigma\cdot\gamma$,
\(\langle S_\Sigma^{(1)}\rangle_{\rm tw}=\mathrm{sgn}(s-\tfrac12)\) for $s\ne\tfrac12$.
At $T>0$, conservation alone does not pin an expectation value: the Gibbs state sums membrane sectors. At the two commuting fixed points this sum is elementary. Each stabilizer in the membrane has thermal expectation $\tanh\beta$, while the twist supplies the braiding sign, hence
\begin{equation}
\langle S_\Sigma^{(1)}\rangle_{\rm tw}=\mp\,\tanh^{|\Sigma|}\!\beta,\qquad |\Sigma|\sim L^2\quad(s=0,1),
\end{equation}
with the minus (plus) sign at $s=0$ ($s=1$). Since $|\Sigma|\propto L^2$, the bare membrane vanishes in the thermodynamic limit for every $T>0$ at the fixed points. Away from them this area-law suppression follows rigorously wherever the high-temperature cluster expansion converges. At lower temperature, exact finite-size traces and QMC---sign-free in the Hamiltonian sampling, though its membrane estimator carries a measurement-basis sign problem---observe no revival over the accessible window (Sec.~\ref{supp:sm:3d}). Thus, in the analytically controlled and numerically accessible regimes, the weak-Gibbs SPT order contracts onto the $T=0$ axis [Fig.~\ref{fig:3d}(a)], whereas its dual can remain ordered at finite temperature.

The cluster model has additional thermal structure on its self-dual line. At $s=\tfrac12$, $U$ is a genuine $\mathbb Z_2$ symmetry and $F(s,T)=F(1-s,T)$ exactly~\cite{Pivot,PivotModels}. Define the conjugate, $U$-odd operators
\(\mathcal M\coloneqq\partial_sH=\sum_i(X_iB_i-X_i)\) and
\(\tilde{\mathcal M}\coloneqq\partial_s\tilde H=\sum_i(B_i-X_i)\).
At every temperature the Hellmann--Feynman identity gives
\begin{equation}\label{eq:HF}
\frac{\partial F(s,T)}{\partial s}=\big\langle \partial_sH(s)\big\rangle_{\beta},
\end{equation}
which follows by inserting
\(\partial_s e^{-\beta H}=-\textstyle\int_0^\beta d\tau\,e^{-(\beta-\tau)H}(\partial_sH)e^{-\tau H}\)
and using cyclicity of the trace. Since $\mathcal M$ is $U$-odd, $s-\tfrac12$ is its conjugate field. A nonzero two-sided limit of $\partial_sF$ therefore means both spontaneous breaking of $U$ and a kink of the free energy---a first-order transition by definition (Sec.~\ref{supp:sm:1d}). This ordinary duality breaking is distinct from the one-form SPT order just shown to vanish. In one dimension the same order parameter $\langle\mathcal M\rangle/(2L)=e_0'(s)$ instead vanishes \emph{continuously} at $s=\tfrac12$ with no jump or hysteresis [Fig.~\ref{fig:1d}, inset], so $U$ stays unbroken and the self-dual point is continuous---the direct contrast to the first-order walls established below.

At nonzero temperature an extensive family of nearly degenerate link configurations changes the cluster-side structure near the self-dual line. The commuting \emph{link} families $\{X_lB_l\}\cup\{X_l\}$ admit $2^{L^3+2}$ bare simultaneous $+1$ configurations, with entropy density approaching $\ln2/6$ per qubit. This count is not an exact degeneracy of the full Hamiltonian: plaquette terms can split the manifold. It nevertheless supplies the low-energy frozen branch once temperature exceeds its tunneling scale (Sec.~\ref{supp:sm:3d}). QMC finds that its configurational entropy opens a narrow wedge around $s=\tfrac12$, bounded by two ordered--frozen first-order lines [Fig.~\ref{fig:3d}(a)]. Their hysteresis strips shrink and the coexistence signal disappears at the endpoints
\(s=\tfrac12\mp0.030(8)\), \(T=0.27(2)\).

No locally saturated analogue exists for $\tilde H$: incident terms anticommute,
\(\{B_p,X_l\}=0\) for $l\in\partial p$, implying
\(\langle B_p\rangle^2+\langle X_l\rangle^2\le1\).
Consistently, the cluster hysteresis develops a central frozen shelf, while the gauge-side loop converts in one step (Sec.~\ref{supp:sm:3d}).

The shape of the dual dome follows from the gauge-theory description. One link-qubit copy of $\tilde H$ is
\(-(1-s)\sum_pB_p-s\sum_lX_l\), the three-dimensional $\mathbb Z_2$ gauge theory in transverse field $h^x\coloneqq s/(1-s)$. Its low-temperature deconfined phase has perimeter-law Wilson loops and broken one-form symmetry. At $h^x=0$, Wegner duality gives~\cite{Wegner}
\begin{equation}\label{eq:Tc-gauge}
e^{-2/T_c(0)}=\tanh K_c^{\rm Ising}\ \Rightarrow\ T_c(0)\approx1.3133,
\end{equation}
where $K_c^{\rm Ising}\approx0.2216$~\cite{FerrenbergXuLandau}. Equation~(\ref{eq:Tc-gauge}) fixes the exact $s=0$ height of the dome. Turning on $h^x$ lowers $T_c$ and eventually destroys deconfinement at the self-dual $T=0$ point $h^x=1$, or $s=\tfrac12$~\cite{ReissSchmidt}.

The order of this boundary changes along the dome. At finite temperature the imaginary-time extent is finite, so wherever the boundary is continuous the transition dimensionally reduces to a three-dimensional classical $\mathbb Z_2$ gauge theory and has three-dimensional Ising criticality. At the $T=0$ endpoint, however, imaginary time becomes infinite and the problem is the self-dual four-dimensional $\mathbb Z_2$ gauge theory, whose transition is weakly first order~\cite{ReissSchmidt,BalianDrouffeItzykson,CreutzJacobsRebbi}. The crossing of two gapped branch free energies with unequal slopes survives weak thermal corrections, so a first-order segment rises from the endpoint and meets the continuous arm at a tricritical point [Fig.~\ref{fig:3d}(b)]. Wegner self-duality pins coexistence exactly at $s=\tfrac12$ in the $P_+$ sector. Including all charge sectors shifts the crossing. A dilute-defect expansion gives only the asymptotic result
\(s_c(T)-\tfrac12=O(T e^{-\Delta_q/T})\) as $T\to0$, for a positive charge-defect gap $\Delta_q$; it fixes neither the prefactor nor the temperature at which this regime becomes visible (Sec.~\ref{supp:sm:3d}).

QMC locates the continuous arm from finite-size scaling of the specific-heat peak and the first-order segment from two-branch coexistence. Over $0.04\le T\le0.22$ the metastability-strip midpoint moves from $s_c\simeq0.540$, near the finite-size endpoint, down to $\simeq0.510$ at the lowest simulated temperature $T=0.04$, consistent within the metastability interval with the exact endpoint $(\tfrac12,0)$; the interval widens only mildly at low $T$. Near $s=0$ the arm gives
\(T_c(s)=T_c(0)[1-s-0.21(2)s^2+O(s^3)]\),
while the arm crosses the self-dual line continuously at
\(T_c(\tfrac12)=0.480(6)\). The low-temperature first-order segment is resolved down to a finite-size metastability endpoint near
\(s\simeq0.542\), \(0.23<T<0.24\), where the resolvable metastable branch splitting disappears; pinning the thermodynamic tricritical point would require multicanonical sampling (Sec.~\ref{supp:sm:3d}). Thus the dual has a genuine finite-temperature deconfined dome rather than a crossover inherited from finite size.

\textit{Role of dimensionality.---}The projected identity is dimension-blind,
but the thermally populated defects are not. In the
one-dimensional dual Ising chains, finite-energy domain walls proliferate at
every $T>0$. In the three-dimensional gauge theory the magnetic defects are
loops, and a positive line tension suppresses them below $T_c$. The cluster
diagnostic, meanwhile, remains a product of thermally fluctuating local
stabilizers. Its decay is exact in one dimension and at the
three-dimensional fixed points, rigorous at sufficiently high temperature,
and shows no revival over the accessible numerical window. This difference in
defect dimensionality makes the one-dimensional agreement accidental and the
three-dimensional disagreement robust.

\textit{Conclusion.---}Non-invertible KT duality relates an all-plus projected
thermal theory; by itself it does not determine the thermodynamics of the Gibbs state. The exact
endpoint and the two QMC phase diagrams show that sectors in the kernel can
alter a bulk singularity when their entropy is extensive; the
one-dimensional agreement occurs only because neither model can
sustain thermal order. More generally, a thermodynamic use of a
non-invertible duality requires independent control of the discarded sectors,
for example a proof that their contribution to $\ln Z$ is subextensive.
Without such control, the fusion and intertwining relations determine
projected data but do not determine physical phase boundaries. The same issue
should arise in other lattice realizations of Kramers--Wannier and
fusion-category dualities~\cite{Kaidi:2022cpf,Bhardwaj:2022maz,Bhardwaj:2022kot,Diatlyk:2023fwf,Seifnashri:2025fgd}.

\textit{Acknowledgments.---}We thank Takamasa Ando, Ryohei Kobayashi, Linhao Li, Ken Shiozaki, and Ryan Thorngren for helpful discussions.

\nocite{BalianDrouffeItzyksonII,FradkinSusskind,KPSDV,KoteckyPreiss,FredenhagenMarcu,PMRQMC,ParaToric,WuDengProkofev}
\bibliography{refs}

\clearpage
\onecolumngrid
\begin{center}
{\large\bfseries Supplemental Material}\\[4pt]
{\large\bfseries Distinct finite-temperature phase diagrams of non-invertible Kennedy--Tasaki duals}\\[8pt]
Weiguang Cao and Haruki Watanabe
\end{center}
\vspace{1.2em}
\twocolumngrid

\setcounter{secnumdepth}{3}
\setcounter{section}{0}
\setcounter{equation}{0}
\setcounter{figure}{0}
\setcounter{table}{0}
\renewcommand{\thesection}{S\arabic{section}}
\renewcommand{\theequation}{S\arabic{equation}}
\renewcommand{\thefigure}{S\arabic{figure}}
\renewcommand{\thetable}{S\arabic{table}}
\renewcommand{\theHsection}{S\arabic{section}}
\renewcommand{\theHequation}{S\arabic{equation}}
\renewcommand{\theHfigure}{S\arabic{figure}}
\renewcommand{\theHtable}{S\arabic{table}}

This Supplemental Material supplies the derivations and numerical details underlying the Letter. Section~\ref{supp:sm:1d} gives the exact one-dimensional free-fermion solution, including the boundary-sector bookkeeping needed for the twisted symmetry response, the dual magnetization, and the self-dual point. Section~\ref{supp:sm:3d} treats the one-form twist, the dual deconfinement boundary, the three-dimensional self-dual point, and the QMC phase diagrams. The setup below restates the models and notation so that the discussion is self-contained.

We begin with the partial-isometry property used throughout. Let $N$ denote the total number of qubits, so $N=2L$ for the one-dimensional ring and $N=6L^3$ for the three-dimensional torus. The bilinear representation of the Letter has $a_j(\sigma)\coloneqq(M\sigma)_j$ and $M=M^\top$, and therefore
\begin{equation}
\begin{aligned}
(D^\dagger D)_{\{\sigma'\}\{\sigma''\}}&=\mathcal N^{-2}\!\sum_{\{\sigma\}}(-1)^{\sigma\cdot M(\sigma'+\sigma'')}\\
&=\mathcal N^{-2}\,2^{N}\,\delta_{\sigma'+\sigma''\in\ker M}
\end{aligned},
\end{equation}
Thus $D^\dagger D=\mathcal N^{-2}2^{N}\sum_{\bm n\in\ker M}X^{\bm n}=P_+$ precisely when $\mathcal N^2=2^{N}|\ker M|$; all exponents are understood modulo $2$. The symmetry of the bilinear kernel $(-1)^{\sigma\cdot M\sigma'}$ under $\sigma\leftrightarrow\sigma'$ also gives $D=D^\dagger$, and hence $DD^\dagger=D^\dagger D=P_+$. Conjugation by $U$ transfers both relations to $\tilde D\coloneqq UDU^\dagger$, establishing $\tilde D^\dagger\tilde D=\tilde D\tilde D^\dagger=P_+$. Multiplying $\tilde D\,H(s)=\tilde H(s)\,\tilde D$ on the right by $\tilde D^\dagger$ and taking the trace gives the projected identity used in the Letter. The one-form case follows by the same algebra, with $\ker M$ generated by the local one-form generators and representatives of the noncontractible membrane classes.

\section{Setup: models, lattices, and ensembles}
\label{supp:sm:setup}
We collect the models, lattices, maps, and ensembles used below. Figure~\ref{supp:fig:twist-geom} illustrates the geometry of the three-dimensional twist.

We first fix the models. The three fixed-point Hamiltonians are $H_{\rm SPT}\coloneqq-\sum_iX_iB_i$, $H_{\rm para}\coloneqq-\sum_iX_i$, and $H_{\rm SSB}\coloneqq-\sum_iB_i$, with $B_i$ a finite product of neighboring $Z$-operators (a local interaction term of $H_{\rm SSB}$). The two interpolations are $H(s)\coloneqq(1-s)H_{\rm SPT}+sH_{\rm para}$ (cluster side) and $\tilde H(s)\coloneqq(1-s)H_{\rm SSB}+sH_{\rm para}$ (dual side), $s\in[0,1]$. All couplings are set to unity, so temperature $T\coloneqq1/\beta$ is measured in units of the field.

In one dimension, a ring of $L$ vertices and $L$ links carries one qubit each ($N=2L$). The dressings are $B_v\coloneqq\prod_{l\ni v}Z_l$ and $B_l\coloneqq\prod_{v\in\partial l}Z_v$ ($\partial l$ the two endpoints of link $l$, $l\ni v$ the two links at vertex $v$), and the protecting symmetry is the zero-form $\mathbb Z_2\times\mathbb Z_2$ generated by $S^{(0)}_{\rm vtx}\coloneqq\prod_vX_v$ and $S^{(0)}_{\rm lnk}\coloneqq\prod_lX_l$. Here $H(s)$ is the cluster chain and $\tilde H(s)$ two decoupled transverse-field Ising chains.

In three dimensions, the $3$-torus of linear size $L$ carries one qubit on each of the $N_l\coloneqq3L^3$ links and each of the $3L^3$ plaquettes ($N=6L^3$). The dressings are $B_l\coloneqq\prod_{p\ni l}Z_p$ and $B_p\coloneqq\prod_{l\in\partial p}Z_l$ ($p\ni l$ the plaquettes containing link $l$, $\partial p$ the four links of plaquette $p$), and the protecting symmetry is the one-form $\mathbb Z_2^{(1)}\times\mathbb Z_2^{(1)}$ generated on cube surfaces and at vertices by $S^{(1)}_c\coloneqq\prod_{p\in\partial c}X_p$ and $S^{(1)}_v\coloneqq\prod_{l\ni v}X_l$ (the cube and vertex $B$-dressings cancelling pairwise). Here $H(s)$ is the RBH cluster interpolation and $\tilde H(s)$ two decoupled transverse-field $\mathbb Z_2$ gauge theories.

Turning to the maps and projectors, a finite-depth diagonal entangler $U$ (a $CZ$ circuit) obeys $UX_iU^\dagger=X_iB_i$, so $H_{\rm SPT}=UH_{\rm para}U^\dagger$; the Kramers--Wannier gauging $D$ obeys $DX_i=B_iD$ and $DB_i=X_iD$; and the Kennedy--Tasaki map is $\tilde D\coloneqq UDU^\dagger$. On a closed manifold $\tilde D$ is a partial isometry onto the all-plus full-fusion sector, $\tilde D^\dagger\tilde D=\tilde D\tilde D^\dagger=P_+$. All twisted-membrane and phase-diagram statements below refer to the Gibbs ensemble unless $P_+$ is written explicitly.

\section{One dimension: exact solution and derivations}
\label{supp:sm:1d}

\subsection{Exact free-fermion solution: conventions, boundary conditions, and parity}
\label{supp:sm:ff}

All finite-temperature statements about $H(s)$ below follow from an exact free-fermion solution. We spell out the boundary bookkeeping because it controls the twist formula, the dual chains, and the self-dual point.

\subsubsection{The unified chain}
Relabel the $2L$ qubits by a single index $j=0,\dots,2L-1$ around the ring, vertices on even $j$ and links on odd $j$. Every dressing takes the uniform form $B_j\coloneqq Z_{j-1}Z_{j+1}$ (indices mod $2L$), and
\begin{equation}\label{supp:eq:H1d-unified}
H(s)=-\sum_{j=0}^{2L-1}\Big[(1-s)\,Z_{j-1}X_jZ_{j+1}+s\,X_j\Big].
\end{equation}
The one-site translation $j\to j+1$ is a symmetry of Eq.~(\ref{supp:eq:H1d-unified}) that exchanges the vertex and link sublattices; consequently every sublattice-resolved quantity computed below is the same for both sublattices.

\subsubsection{Jordan--Wigner transformation and the boundary terms}
Define $2L$ pairs of Majorana operators ($Y_j\coloneqq iX_jZ_j$),
\begin{equation}\label{supp:eq:JW}
\gamma_{2j}\coloneqq\Big(\prod_{k<j}X_k\Big)Z_j,\qquad
\gamma_{2j+1}\coloneqq-\Big(\prod_{k<j}X_k\Big)Y_j,
\end{equation}
which are Hermitian and obey $\{\gamma_a,\gamma_b\}=2\delta_{ab}$. Direct multiplication gives the identities needed below. On any site the string commutes with $Z_j,Y_j$ and squares to one; using $Z_jY_j=-iX_j$,
\begin{equation}\label{supp:eq:Xj-maj}
X_j=-i\,\gamma_{2j}\gamma_{2j+1}.
\end{equation}
For a \emph{bulk} stabilizer ($1\le j\le 2L-2$), the two strings overlap on $k<j-1$ and cancel there, leaving $X_{j-1}X_j$. Using $Y_{j-1}X_{j-1}=-iZ_{j-1}$,
\begin{align}
\gamma_{2j-1}\gamma_{2j+2}
&=-Y_{j-1}\,X_{j-1}X_j\,Z_{j+1}\nonumber\\
&=i\,Z_{j-1}X_jZ_{j+1},\nonumber\\
\Rightarrow\;
Z_{j-1}X_jZ_{j+1}&=-i\,\gamma_{2j-1}\gamma_{2j+2}.
\label{supp:eq:bulk-maj}
\end{align}
The two stabilizers that wrap the ring require separate treatment. They are \emph{not} of the form~(\ref{supp:eq:bulk-maj}), because their Jordan--Wigner strings complete to the total fermion parity
\begin{equation}
P\coloneqq\prod_{j=0}^{2L-1}X_j=\prod_{j}\big(-i\gamma_{2j}\gamma_{2j+1}\big),
\end{equation}
and an explicit multiplication (using $X_0Z_0=-iY_0$, $X_jY_j=iZ_j$) gives
\begin{equation}\label{supp:eq:boundary-maj}
\begin{aligned}
Z_{2L-2}X_{2L-1}Z_{0}&=+i\,P\,\gamma_{4L-3}\gamma_{0},\\
Z_{2L-1}X_{0}Z_{1}&=+i\,P\,\gamma_{4L-1}\gamma_{2}.
\end{aligned}
\end{equation}
Relative to the bulk form~(\ref{supp:eq:bulk-maj}) continued around the ring, each wrapping bond carries the extra factor $(-P)$. We verified Eqs.~(\ref{supp:eq:Xj-maj})--(\ref{supp:eq:boundary-maj}) as exact operator identities numerically (deviation $0$ in operator norm on the $2^{2L}$-dimensional Hilbert space, for $L\le8$).

\subsubsection{Parity--boundary-condition locking}
Since $[P,H(s)]=0$, the Hilbert space splits into sectors $P=p=\pm1$, and within a sector the wrapping factor $(-P)$ is the number $(-p)$. Defining the antiperiodic continuation $\gamma_{a+4L}\coloneqq-p\,\gamma_a$, the Hamiltonian takes the uniform quadratic form
\begin{equation}\label{supp:eq:H-majorana}
H(s)\big|_{P=p}=i\sum_{j=0}^{2L-1}\Big[s\,\gamma_{2j}\gamma_{2j+1}+(1-s)\,\gamma_{2j-1}\gamma_{2j+2}\Big],
\end{equation}
with \emph{antiperiodic} (Neveu--Schwarz, NS) Majoranas for $p=+1$ and \emph{periodic} (Ramond, R) Majoranas for $p=-1$. Parity therefore locks the boundary condition and selects the momentum lattice: $k\in\{(2n+1)\pi/L\}$ in the NS sector and $k\in\{2\pi n/L\}$ in the R sector, with $n=0,\dots,L-1$. Neither sector may be discarded; the thermal traces below sum over both with projectors $\tfrac12(1+pP)$.

\subsubsection{Two decoupled chains and their diagonalization}
The couplings in Eq.~(\ref{supp:eq:H-majorana}) connect Majorana indices $a\to a+1$ ($a$ even) and $a\to a+3$ ($a$ odd); both preserve $a\bmod 4\in\{0,1\}$ versus $\{2,3\}$. The model therefore decouples into two independent Majorana chains---chain $V$, built from the even-site Majoranas $\{\gamma_{4m},\gamma_{4m+1}\}$, and chain $\Lambda$, from the odd-site Majoranas $\{\gamma_{4m+2},\gamma_{4m+3}\}$, $m=0,\dots,L-1$ (the Majoranas of site $j=2m$ and $j=2m+1$, respectively). Chain $\Lambda$ collects the fields $X_j$ on odd $j$ ($s$) and the stabilizers centered on even $j$ ($1-s$); chain $V$ collects the fields on even $j$ and the stabilizers centered on odd $j$. Their fermion parities are exactly the two symmetry generators,
\begin{equation}\label{supp:eq:parity-id}
\begin{aligned}
P_V&\coloneqq\prod_{j\,{\rm even}} X_j=S^{(0)}_{\rm vtx},\qquad
P_\Lambda\coloneqq\prod_{j\,{\rm odd}} X_j=S^{(0)}_{\rm lnk},\\
P&=P_VP_\Lambda,
\end{aligned}
\end{equation}
and the two chains are copies of the same $L$-site chain (they are exchanged by the one-site translation). Writing $a_m\coloneqq\gamma_{4m+2}$, $b_m\coloneqq\gamma_{4m+3}$ and $c_m\coloneqq\tfrac12(a_m+ib_m)$, chain $\Lambda$ is the transverse-field Ising chain in its fermionic form; Fourier--Bogoliubov diagonalization gives, in either momentum sector,
\begin{equation}\label{supp:eq:bogoliubov}
H_\Lambda=\sum_{k}\varepsilon_k(s)\Big(\eta^\dagger_k\eta_k-\tfrac12\Big),\qquad
\frac{\varepsilon_k(s)}{2}\coloneqq\big|(1-s)e^{ik}-s\big|,
\end{equation}
with Bogoliubov angle $\theta_k$ defined by
\begin{equation}\label{supp:eq:bogangle}
\sin\theta_k\coloneqq\frac{2(1-s)\sin k}{\varepsilon_k},\quad
\cos\theta_k\coloneqq\frac{2\big[s-(1-s)\cos k\big]}{\varepsilon_k}.
\end{equation}
In the R sector the $k=0$ mode is unpaired and enters as a single fermion of \emph{signed} energy $\tilde\varepsilon_0\coloneqq2(2s-1)$ (and, for $L$ even, $k=\pi$ is unpaired with $\tilde\varepsilon_\pi\coloneqq+2$, always positive); all other modes come in $\pm k$ pairs with $\varepsilon_k>0$.

For a single $L$-site chain we shall need four traces: the ordinary and parity-inserted partition functions in each sector $\sigma\in\{{\rm NS},{\rm R}\}$,
\begin{equation}\label{supp:eq:fourZ}
\begin{aligned}
Z^{\sigma}(\beta)&\coloneqq\mathrm{Tr}_\sigma\,e^{-\beta H}=\prod_{k\in\sigma}2\cosh\frac{\beta\varepsilon_k}{2},\\
\tilde Z^{\sigma}(\beta)&\coloneqq\mathrm{Tr}_\sigma\big(S^{(0)}_{\rm lnk}\,e^{-\beta H}\big)=\prod_{k\in\sigma}2\sinh\frac{\beta\tilde\varepsilon_k}{2},
\end{aligned}
\end{equation}
where the inserted operator is the chain's fermion parity $S^{(0)}_{\rm lnk}=(-1)^{\sum_k\eta_k^\dagger\eta_k}$---the symmetry generator $P_\Lambda$ of Eq.~(\ref{supp:eq:parity-id}), one factor of the global parity $P=P_VP_\Lambda$, not the symmetric-sector projector $P_+$ of the Letter---and $\tilde\varepsilon_k\coloneqq\varepsilon_k$ except for the signed R-sector zero mode above. The elementary identity behind the second line is the single-mode average $\mathrm{Tr}[(-1)^{\hat n}e^{-\beta\varepsilon\hat n}]/\mathrm{Tr}[e^{-\beta\varepsilon\hat n}]=\tanh(\beta\varepsilon/2)$, which is negative for the signed mode when $s<\tfrac12$. With these conventions the sector decomposition of the full cluster chain,
\begin{equation}\label{supp:eq:Zdecomp}
\mathrm{Tr}\,e^{-\beta H(s)}
=\tfrac12\Big[(Z^{\rm NS})^2+(\tilde Z^{\rm NS})^2\Big]+\tfrac12\Big[(Z^{\rm R})^2-(\tilde Z^{\rm R})^2\Big],
\end{equation}
reproduces exact diagonalization to machine precision for every accessible size ($2L\le16$ qubits). The same check holds for all sector formulas below. The two factors in each product are the $V$ and $\Lambda$ chains, while the signs implement $\tfrac12\sum_p(1+pP)$ with the boundary conditions locked to $p$. The physical ground state lies in the $p=+1$ (NS$\otimes$NS) sector and is the Bogoliubov vacuum of both chains, with total parity $+1$.

\subsection{Derivation of the twisted symmetry-flux formula of the Letter}

The cluster SPT is diagnosed by a symmetry-flux response, threading an $S_{\rm vtx}^{(0)}$ flux through the ring (a twisted boundary condition for the vertex symmetry) and measuring $S_{\rm lnk}^{(0)}$. The twist inserts a branch cut on one bond. The single stabilizer straddling the cut is centered on an odd site $j_0$ and carries $B_{j_0}=Z_{j_0-1}Z_{j_0+1}$, so the crossing $Z$ is conjugated to $-Z$ and exactly that stabilizer flips, $X_{j_0}B_{j_0}\to -X_{j_0}B_{j_0}$. The twisted ground state is $Z_{j_0}\lvert C\rangle$, on which $S_{\rm lnk}^{(0)}=-1$. In the paramagnet every term is on-site, the twist is inert, and $S_{\rm lnk}^{(0)}=+1$.

At finite temperature $H_{\rm SPT}$ and $H_{\rm para}$ are each a sum of $2L$ commuting $\pm1$ generators with $\langle X_iB_i\rangle_\beta=\langle X_i\rangle_\beta=\tanh\beta$, so the product $S_{\rm lnk}^{(0)}$ factorizes,
\begin{equation}\label{supp:eq:Slnk4}
\begin{aligned}
\langle S_{\rm lnk}^{(0)}\rangle^{\rm SPT}_{\rm PBC}
&=\langle S_{\rm lnk}^{(0)}\rangle^{\rm para}_{\rm PBC}
=\langle S_{\rm lnk}^{(0)}\rangle^{\rm para}_{\rm tw}
=+\tanh^{L}\!\beta ,\\
\langle S_{\rm lnk}^{(0)}\rangle^{\rm SPT}_{\rm tw}
&=-\tanh^{L}\!\beta .
\end{aligned}
\end{equation}
All four have the same magnitude. The SPT differs from the trivial phase through the \emph{sign} of the twisted expectation, which represents the type-II class of $H^2(\mathbb Z_2\times\mathbb Z_2,U(1))=\mathbb Z_2$~\cite{PT}. In the standard labelling for product groups, a type-I class involves a single factor and is trivial here because $H^2(\mathbb Z_2,U(1))=0$. The type-II class instead pairs the two factors: a flux of the vertex $\mathbb Z_2$ binds a charge of the link $\mathbb Z_2$, equivalently producing an anticommuting projective action of the two symmetry generators at an edge. The common magnitude is
\begin{equation}\label{supp:eq:xi-1d-spt}
\begin{aligned}
\big|\langle S_{\rm lnk}^{(0)}\rangle\big|&=\tanh^{L}\!\beta=e^{-L/\xi},\\
\xi&\coloneqq\frac{1}{\ln\coth\beta}\simeq\tfrac12\,e^{2/T},
\end{aligned}
\end{equation}
and decays with a correlation length finite at every $T>0$. Taking $L\to\infty$ first therefore makes the order parameter vanish. The bulk thermodynamics is insensitive to the SPT distinction: $H_{\rm SPT}=UH_{\rm para}U^\dagger$ gives $Z_{\rm SPT}=Z_{\rm para}=(2\cosh\beta)^{2L}$, which is analytic in $\beta$, while the twisted sign in Eq.~(\ref{supp:eq:Slnk4}) distinguishes the two phases.

Along the interpolation~(\ref{supp:eq:H1d-unified}), $S_{\rm lnk}^{(0)}$ stays conserved. At $T=0$ it is pinned to a symmetry eigenvalue,
\begin{equation}\label{supp:eq:Sofs-T0}
\langle S_{\rm lnk}^{(0)}\rangle_{\rm PBC}=+1,\qquad
\langle S_{\rm lnk}^{(0)}\rangle_{\rm tw}=\mathrm{sgn}\!\big(s-\tfrac12\big)\qquad(T=0),
\end{equation}
a quantized invariant equal to $-1$ across the SPT phase and $+1$ across the trivial phase, jumping at $s=\tfrac12$. The sector decomposition of Sec.~\ref{supp:sm:ff} determines its finite-temperature rounding exactly.

In the Majorana language the twist is a one-bond defect. It flips the single odd-centered stabilizer $X_{j_0}B_{j_0}$, which by Eq.~(\ref{supp:eq:bulk-maj}) is one bond of chain $V$. Flipping the signs of the chain-$V$ Majoranas on the arc between the defect and the boundary bond is an orthogonal transformation that restores the bulk form and instead reverses the sign of chain $V$'s wrapping bond. Within the sector $P=p$ the boundary conditions of the two chains are therefore no longer equal. Chain $V$ is twisted to R (for $p=+1$) or NS (for $p=-1$), while chain $\Lambda$ retains NS ($p=+1$) or R ($p=-1$). The measured operator is $S^{(0)}_{\rm lnk}=P_\Lambda$, Eq.~(\ref{supp:eq:parity-id}). Assembling the traces with the projectors $\tfrac12(1+pP)$, $P=P_VP_\Lambda$, and using $\mathrm{Tr}(P_VP_\Lambda\cdot P_\Lambda\,e^{-\beta H})=\tilde Z_V Z_\Lambda$,
\begin{equation}\label{supp:eq:sector-tw}
\begin{aligned}
\mathrm{Tr}\,e^{-\beta H_{\rm tw}}
&=\tfrac12\big[Z^{\rm R}Z^{\rm NS}+\tilde Z^{\rm R}\tilde Z^{\rm NS}\big]\\
&\quad+\tfrac12\big[Z^{\rm NS}Z^{\rm R}-\tilde Z^{\rm NS}\tilde Z^{\rm R}\big]
=Z^{\rm R}\,Z^{\rm NS},\\
\mathrm{Tr}\big(S^{(0)}_{\rm lnk}e^{-\beta H_{\rm tw}}\big)
&=\tfrac12\big[Z^{\rm R}\tilde Z^{\rm NS}+\tilde Z^{\rm R}Z^{\rm NS}\big]\\
&\quad+\tfrac12\big[Z^{\rm NS}\tilde Z^{\rm R}-\tilde Z^{\rm NS}Z^{\rm R}\big]
=\tilde Z^{\rm R}\,Z^{\rm NS},
\end{aligned}
\end{equation}
where in each bracket the first (second) term is the $p=+1$ ($p=-1$) sector with the boundary conditions just listed, and the cancellations use the fact that chains $V$ and $\Lambda$ are identical copies. Both lines of Eq.~(\ref{supp:eq:sector-tw}) reproduce exact diagonalization to machine precision. Every parity-inserted factor of chain $V$ has cancelled, so no convention adopted for it can affect the result; in fact either admissible arc contains an \emph{even} number of chain-$V$ Majoranas [$2(m_0+1)$ or $2(L-1-m_0)$ for the defect at $j_0=2m_0+1$], so the transformation preserves $P_V$ outright. The ratio is a single parity-inserted R-sector average of the link chain,
\begin{equation}
\langle S_{\rm lnk}^{(0)}\rangle_{\rm tw}
=\frac{\tilde Z^{\rm R}}{Z^{\rm R}}
=\prod_{k\in{\rm R}}\tanh\frac{\beta\tilde\varepsilon_k(s)}{2},
\end{equation}
which factorizes over the periodic momenta $k_n\coloneqq2\pi n/L$, the unpaired $k=0$ mode carrying the signed energy $\tilde\varepsilon_0=2(2s-1)$ of Sec.~\ref{supp:sm:ff} and hence the signed factor $\tanh[\beta(2s-1)]$:
\begin{equation}\label{supp:eq:Sofs-tw}
\langle S_{\rm lnk}^{(0)}\rangle_{\rm tw}=\tanh\!\big[\beta(2s-1)\big]\prod_{n=1}^{L-1}\tanh\frac{\beta\,\varepsilon_{k_n}(s)}{2},
\end{equation}
with $\varepsilon_k(s)$ the dispersion of Eq.~(\ref{supp:eq:bogoliubov}). The signed $k=0$ prefactor carries the jump of Eq.~(\ref{supp:eq:Sofs-T0}) as $T\to0$. For fixed $s\neq\tfrac12$, every factor tends to $\pm1$ as $\beta\to\infty$, and the product tends to $\mathrm{sgn}(2s-1)$.

The limits at the critical point require care. Quantization is obtained by taking $s\to\tfrac12^{\pm}$ before $\beta\to\infty$. At $s=\tfrac12$ exactly, the signed zero mode $\tilde\varepsilon_0=2(2s-1)$ vanishes and the prefactor $\tanh[\beta(2s-1)]$ is zero at every finite $\beta$, so $\langle S_{\rm lnk}^{(0)}\rangle_{\rm tw}=0$. The noncommuting limits reflect the selection of one of the two ground states reached from $s\lessgtr\tfrac12$. The $\coth$-weighted expressions below use the same prescription: their R-sector $k=0$ factor is individually singular and cancels only after the full sector-weighted trace is assembled.

At the fixed point $s=0$, where $\varepsilon_k/2=1$ for all $k$ and $\tilde\varepsilon_0=-2$, Eq.~(\ref{supp:eq:Sofs-tw}) reduces to $-\tanh^L\!\beta$, recovering Eq.~(\ref{supp:eq:Slnk4}). More generally, conservation gives the thermal expectation $\langle S_{\rm lnk}^{(0)}\rangle=\sum_n p_n\,e^{-\beta E_n}/Z$, where $p_n=\pm1$ is the eigenvalue in the $n$-th energy eigenstate. This Boltzmann average reduces to a quantized charge only at $T=0$ and is a smooth reweighting at $T>0$. Accordingly, the curve is smooth in $s$ for every nonzero temperature, and the ordered region contracts onto the $T=0$ axis as $L\to\infty$. Figure~2(a) of the Letter evaluates the exact product~(\ref{supp:eq:Sofs-tw}) at $L=10^6$; at $s=0$ the finite-size lobe ends near $T=0.13$.

\subsection{Derivation of the dual-side magnetization and its finite-temperature decay}

On each sublattice $\tilde H(s)$ is a transverse-field Ising chain, self-dual at $s=\tfrac12$, and the two sublattices are identical. Each breaks its own $\mathbb Z_2$ in the ordered phase, so that the full $\mathbb Z_2\times\mathbb Z_2$ is spontaneously broken. This breaking is captured exactly by the squared magnetization of a single sublattice---say the even (vertex) sites of the unified chain of Sec.~\ref{supp:sm:ff}---namely
\begin{equation}\label{supp:eq:m2-def}
m^2\coloneqq\frac{1}{L^2}\Big\langle\Big(\sum_{j\,{\rm even}} Z_j\Big)^{\!2}\Big\rangle=\frac{1}{L}\sum_{j\,{\rm even}}\langle Z_0Z_j\rangle,
\end{equation}
with $j$ running over the $L$ even sites $0,2,\dots,2L-2$; by translation invariance $\langle Z_0Z_j\rangle$ depends only on the separation $j$. Since the two chains decouple and each is $\mathbb Z_2$-symmetric, the correlators between sites of opposite parity vanish identically, $\langle Z_jZ_{j'}\rangle=\langle Z_j\rangle\langle Z_{j'}\rangle=0$ ($j$ even, $j'$ odd), so restricting to the even sites loses nothing (and including the odd sites in the sum would only halve the normalization). Jordan--Wigner, with the same conventions as Eq.~(\ref{supp:eq:JW}) applied to the $L$-site even sublattice~\cite{Pfeuty,BarouchMcCoy}, maps it to a free-fermion chain with the dispersion $\varepsilon_k(s)$ of Eq.~(\ref{supp:eq:bogoliubov}), and on a periodic ring $\langle Z_0Z_j\rangle_L$ is the exact parity-weighted sum over the Neveu--Schwarz and Ramond sectors---the boundary bookkeeping of Sec.~\ref{supp:sm:ff}, with the wrapped Ising bond $Z_{2L-2}Z_0$ carrying the chain parity exactly as in Eq.~(\ref{supp:eq:boundary-maj}). In closed form---the dual-side analogue of Eq.~(\ref{supp:eq:Sofs-tw})---it reads, with the four traces of Eq.~(\ref{supp:eq:fourZ}),
\begin{equation}\label{supp:eq:ZZ-exact}
\langle Z_0Z_j\rangle_L=\frac{Z^{\rm NS}\mathsf D^{\rm NS}+\tilde Z^{\rm NS}\tilde{\mathsf D}^{\rm NS}+Z^{\rm R}\mathsf D^{\rm R}-\tilde Z^{\rm R}\tilde{\mathsf D}^{\rm R}}{Z^{\rm NS}+\tilde Z^{\rm NS}+Z^{\rm R}-\tilde Z^{\rm R}},
\end{equation}
where $\mathsf D^{\sigma}\coloneqq\det_{1\le a,b\le j/2}(G^{\sigma}_{a-b+1})$ and $\tilde{\mathsf D}^{\sigma}\coloneqq\det_{1\le a,b\le j/2}(\tilde G^{\sigma}_{a-b+1})$ are Toeplitz determinants built on the discrete momenta of sector $\sigma\in\{{\rm NS},{\rm R}\}$. The subscript $L$ marks a finite-ring correlator and is dropped after taking the thermodynamic limit. The matrix elements are
\begin{equation}\label{supp:eq:Gn-exact}
G^{\sigma}_n\coloneqq\frac1L\sum_{k\in\sigma}e^{-ink}\,
\frac{(1-s)e^{ik}-s}{\tilde\varepsilon_k/2}\,
\tanh\frac{\beta\tilde\varepsilon_k}{2},
\end{equation}
and $\tilde G^{\sigma}_n$ is obtained by replacing $\tanh$ with $\coth$. Here $k=(2n+1)\pi/L$ in the NS sector and $k=2\pi n/L$ in the R sector, with $n=0,\dots,L-1$. The signed convention for $\tilde\varepsilon_k$ is that of Eq.~(\ref{supp:eq:fourZ}), so each ordinary trace carries the $\tanh$ occupation weight and each parity-inserted trace the $\coth$ weight of its sector.

Equation~(\ref{supp:eq:ZZ-exact}) reproduces exact diagonalization to machine precision and is computationally comparable to Eq.~(\ref{supp:eq:Sofs-tw}). A single triangular factorization gives all nested minors $j/2=1,\dots,L/2$ of each Toeplitz matrix; separations beyond $L/2$ follow from $\langle Z_0Z_j\rangle=\langle Z_0Z_{2L-j}\rangle$. Substitution into Eq.~(\ref{supp:eq:m2-def}) then evaluates $m^2(s,T)$ for rings up to $L\sim10^3$ in milliseconds per point.

At fixed $T>0$, the parity-inserted terms are exponentially suppressed as $L\to\infty$; the R-sector ratio $\tilde Z^{\rm R}/Z^{\rm R}$ is precisely the twist product of Eq.~(\ref{supp:eq:Sofs-tw}). The momentum sums become integrals, and Eq.~(\ref{supp:eq:ZZ-exact}) reduces to the bulk Toeplitz determinant
\begin{equation}\label{supp:eq:toeplitz1d}
\begin{aligned}
\langle Z_0 Z_j\rangle&=\det_{1\le a,b\le j/2}\big(G_{a-b+1}\big)\qquad(j\ {\rm even}),\\
G_n&\coloneqq\frac{1}{2\pi}\int_0^{2\pi}\!\!dk\;e^{-ink}\,\frac{(1-s)e^{ik}-s}{\big|(1-s)e^{ik}-s\big|}\,\tanh\frac{\beta\varepsilon_k}{2};
\end{aligned}
\end{equation}
At $T=0$, the occupations saturate and Szeg\H{o}'s theorem gives the spontaneous magnetization
\begin{equation}\label{supp:eq:m2T0}
m^2\;\xrightarrow{\,T=0\,}\;
\begin{cases}
m_0^2\coloneqq\big(1-g^2\big)^{1/4}, & s<\tfrac12,\\[2pt]
0, & s>\tfrac12,
\end{cases}
\quad g\coloneqq\frac{s}{1-s},
\end{equation}
the order parameter \emph{vanishing continuously} as $s\to\tfrac12^-$ (since $g\to1$, so $m_0^2=(1-g^2)^{1/4}\to0$), not jumping. This continuous vanishing of the SSB order is distinct from the discontinuous sign of the twisted SPT invariant when the critical point is approached from the two gapped phases.

At any $T>0$ the modulus of the Toeplitz symbol, $\tanh(\beta\varepsilon_k/2)<1$, forces exponential decay $\langle Z_0Z_j\rangle\sim A(s,T)\,e^{-j/(2\xi)}$ by the strong Szeg\H{o} theorem ($j/2$ is the number of sublattice steps; the prefactor $A(s,T)$ is a correlation amplitude, not a spontaneous magnetization, which vanishes at every $T>0$), with
\begin{equation}\label{supp:eq:xi1d}
\xi^{-1}\coloneqq-\frac{1}{2\pi}\int_0^{2\pi}\!dk\,\ln\tanh\frac{\beta\varepsilon_k}{2}>0\qquad(s<\tfrac12),
\end{equation}
finite for every $T>0$ and diverging only as $T\to0$. On the ordered side $s<\tfrac12$, the same $\xi$ governs the cluster twist. For $s>\tfrac12$, Eq.~(\ref{supp:eq:xi1d}) continues to govern the twist product, while the dual paramagnet has a distinct, finite zero-temperature quantum correlation length.

Indeed, the logarithm of the product in Eq.~(\ref{supp:eq:Sofs-tw}) is the R-momentum Riemann sum
\(\ln\prod_{n=1}^{L-1}\tanh(\beta\varepsilon_{k_n}/2)=-L/\xi+O(1)\),
where the $O(1)$ term comes from excluding $k=0$. Thus the twisted flux decays with system \emph{size} at the same rate as the ordered-side dual correlator decays with \emph{distance}. For example, at $s=0.3$ and $\beta=1$, both give $\xi^{-1}=0.523026$. This common correlation length is the quantitative finite-temperature content of the KT exchange. Since finite $\xi$ makes $\chi\coloneqq\sum_{j\,{\rm even}}\langle Z_0Z_j\rangle$ converge, on the ring
\begin{equation}\label{supp:eq:m2decay}
m^2(L,T)=\frac1L\sum_{j\,{\rm even}}\langle Z_0Z_j\rangle_L\;\xrightarrow{\,L\gg\xi\,}\;\frac{\chi}{L}\sim\frac1L\;\to\;0 .
\end{equation}
The chain therefore has no finite-temperature symmetry breaking, in agreement with Peierls's domain-wall argument, although the finite-size contraction is exponentially slow at low $T$. Expanding Eq.~(\ref{supp:eq:xi1d}) about the gap $\Delta\coloneqq2(1-2s)$ gives $\xi\sim T^{-1/2}e^{\Delta/T}$ for generic $0<s<\tfrac12$, where the band minimum is quadratic. At $s=0$ the dispersion is flat, and Eq.~(\ref{supp:eq:xi1d}) instead gives the exact result $\xi=1/\ln\coth\beta\sim\tfrac12\,e^{2/T}$. The normalized order parameter consequently has the scaling form $m^2(L,T)/m_0^2=\Phi(L/\xi)$, with $m_0^2$ the zero-temperature value defined above: it appears ordered for $L\ll\xi$ and decays as $\chi/L$ for $L\gg\xi$. The apparent transition therefore approaches the $T=0$ axis only logarithmically, as $\Delta/\ln L$.

Figure~2 of the Letter shows Eq.~(\ref{supp:eq:m2-def}) for $L=10^6$, corresponding to two million qubits. The flux-twist panel is the exact finite-$L$ product~(\ref{supp:eq:Sofs-tw}), whereas the magnetization panel is a controlled, numerically converged large-$L$ evaluation of Eq.~(\ref{supp:eq:ZZ-exact}). The momentum sums in Eq.~(\ref{supp:eq:Gn-exact}) differ from their $L\to\infty$ integrals by $O(e^{-cL})$, with $c\coloneqq|\ln g|$ the width of the analyticity strip of the symbol. The NS and R determinants then coincide, while the four trace weights reduce pairwise to $Z^{\rm R}/Z^{\rm NS}\to1$, $\tilde Z^{\rm NS}/Z^{\rm NS}\to u$, and $\tilde Z^{\rm R}/Z^{\rm R}\to\mp u$ for $s\lessgtr\tfrac12$, where $u\coloneqq e^{-L/\xi}$ and $\xi$ is given by Eq.~(\ref{supp:eq:xi1d}). For $s<\tfrac12$, Eq.~(\ref{supp:eq:ZZ-exact}) becomes
\begin{equation}\label{supp:eq:ZZ-bigL}
\langle Z_0Z_j\rangle_L=\frac{\mathsf D_{j/2}+u\,\tilde{\mathsf D}_{j/2}}{1+u},
\end{equation}
Here $\mathsf D_r$ ($\tilde{\mathsf D}_r$) is the $r$-th Toeplitz determinant built from the common integral coefficients with the $\tanh$ ($\coth$) weight. The parity-inserted term carries the second arm of the ring,
\(u\tilde{\mathsf D}_{j/2}\simeq m_0^2e^{-(L-j/2)/\xi}\).
For $s>\tfrac12$, the parity-inserted weights instead cancel pairwise and leave $\mathsf D_{j/2}+\mathsf D_{L-j/2}$.

For the numerical evaluation, the momentum integrals in Eq.~(\ref{supp:eq:Gn-exact}) use a $2^{15}$-point FFT quadrature, Toeplitz minors beyond $r=600$ are resummed using their exact geometric tails, and the value at $s=\tfrac12$ is the average of those at $s=\tfrac12\pm10^{-6}$. The result is stable to machine precision upon refining the minor cutoff $R$, the FFT grid, or the critical-point offset, and it reproduces Eq.~(\ref{supp:eq:ZZ-exact}) directly wherever that expression is computationally accessible ($L\le4096$). In Fig.~2 of the Letter, $m^2$ saturates $m_0^2=(1-g^2)^{1/4}$ throughout $s<\tfrac12$ at $T=0$. At $T>0$, the apparently ordered region survives only below the logarithmically shrinking crossover: at $s=0$ the dome has half height at $T=0.15$, whereas lowering it to $T=0.1$ would require $L\approx10^9$.

\subsection{The self-dual point: an unbroken duality symmetry}
\label{supp:sm:selfdual}

At $s=\tfrac12$ the reflection $U^\dagger H(s)U=H(1-s)$ closes into a genuine $\mathbb Z_2$ symmetry, $[U,H(\tfrac12)]=0$---the cluster entangler itself becomes a symmetry operator. Spontaneous breaking of this symmetry is tested by the local $U$-odd observable
\begin{equation}\label{supp:eq:Ol}
\mathcal M_j\coloneqq X_jB_j-X_j,\qquad U\,\mathcal M_j\,U^\dagger=-\mathcal M_j,
\end{equation}
since $U$ exchanges $X_j\leftrightarrow X_jB_j$ (it dresses each field and, being an involution with $UB_jU^\dagger=B_j$, undresses the dressed one). By symmetry $\langle \mathcal M_j\rangle=0$ identically in the $U$-symmetric ensemble, so the meaningful diagnostic is the squared order parameter
\begin{equation}\label{supp:eq:m2U}
m_U^2\coloneqq\frac{1}{L^2}\sum_{j,j'\,{\rm odd}}\big\langle \mathcal M_j\mathcal M_{j'}\big\rangle
=\frac{1}{L^2}\Big\langle\Big(\sum_{j\,{\rm odd}}\mathcal M_j\Big)^{\!2}\Big\rangle,
\end{equation}
with $j,j'$ running over the $L$ odd (link) sites (by the one-site translation of Sec.~\ref{supp:sm:ff} the even-site and full-chain versions give identical values). Note that $\mathcal M_j$ is a useful order parameter \emph{only} at $s=\tfrac12$. Away from it $U$ is not a symmetry, $\langle \mathcal M_j\rangle\neq0$, and the correlator saturates trivially to the disconnected piece $\langle \mathcal M_j\rangle^2$.

The ground-state value of Eq.~(\ref{supp:eq:m2U}) can be computed in closed form. Write $\sum_{j\,{\rm odd}}\mathcal M_j=\sum_{j\,{\rm odd}}(y_j-x_j)$ with $x_j\coloneqq X_j$ and $y_j\coloneqq X_jB_j$. By Eqs.~(\ref{supp:eq:Xj-maj}) and~(\ref{supp:eq:bulk-maj}), $x_j$ is a bond of chain $\Lambda$ and $y_j$ a bond of chain $V$; since the ground state is the product of the two chain vacua (Sec.~\ref{supp:sm:ff}), cross-correlators factorize, $\langle x_jy_{j'}\rangle=\langle x_j\rangle\langle y_{j'}\rangle$. The ground state is moreover $U$-invariant ($U\lvert{\rm gs}\rangle=+\lvert{\rm gs}\rangle$, the critical point being unique), so $\langle(\sum y)^2\rangle=\langle(\sum x)^2\rangle$ and $\langle\sum y\rangle=\langle\sum x\rangle$. Hence
\begin{equation}\label{supp:eq:M2var}
L^2\,m_U^2=\Big\langle\Big(\sum_{j\,{\rm odd}}\mathcal M_j\Big)^{\!2}\Big\rangle
=2\,\mathrm{Var}\Big(\sum_{j\,{\rm odd}} X_j\Big).
\end{equation}
Equation~(\ref{supp:eq:M2var}) reduces the calculation to number fluctuations in chain $\Lambda$. In its on-site fermions, $X_{2m+1}=1-2c^\dagger_mc_m$, so $\sum_{j\,{\rm odd}}X_j=L-2\hat N$ and $\mathrm{Var}(\sum_{j\,{\rm odd}} X_j)=4\,\mathrm{Var}(\hat N)$. For the Bogoliubov vacuum $\prod_{k>0}(\cos\tfrac{\theta_k}{2}+\sin\tfrac{\theta_k}{2}\,c^\dagger_kc^\dagger_{-k})\lvert0\rangle$, each $(k,-k)$ pair holds $0$ or $2$ fermions, giving $\mathrm{Var}(\hat N)=\sum_{k>0}\sin^2\theta_k=\tfrac12\sum_{k\in{\rm NS}}\sin^2\theta_k$ (the unpaired $k=\pi$ mode, present for odd $L$, has $\sin\theta_\pi=0$). At the self-dual point Eq.~(\ref{supp:eq:bogangle}) gives $\varepsilon_k=2|\sin(k/2)|$ and
\begin{equation}
\sin\theta_k=\frac{\sin k}{2\sin(k/2)}=\cos\frac k2
\qquad\Big(s=\tfrac12\Big),
\end{equation}
so that, using $\sum_{k\in{\rm NS}}e^{ik}=0$,
\begin{equation}\label{supp:eq:varN}
\mathrm{Var}(\hat N)=\frac12\sum_{k\in{\rm NS}}\cos^2\frac k2
=\frac L4+\frac14\sum_{k\in{\rm NS}}\cos k=\frac L4
\end{equation}
\emph{exactly, at every finite $L$}. Collecting Eqs.~(\ref{supp:eq:M2var})--(\ref{supp:eq:varN}),
\begin{equation}\label{supp:eq:chi2}
\chi_U\coloneqq\sum_{j'\,{\rm odd}}\langle \mathcal M_j\mathcal M_{j'}\rangle=2,
\quad
m_U^2=\frac{\chi_U}{L}=\frac2L\;\xrightarrow{\;L\to\infty\;}\;0
\end{equation}
at $s=\tfrac12$. Exact diagonalization confirms $m_U^2L=2$ to machine precision for every accessible size, with the diagonal part fixed by the operator identity $\mathcal M_j^2=2(1-B_j)$ and the off-diagonal sum by $\sum_{j'\neq j}\langle \mathcal M_j\mathcal M_{j'}\rangle=2\langle B_j\rangle$.

Thus $U$ remains unbroken at the self-dual point. The squared order parameter obeys the susceptibility law $m_U^2=\chi_U/L$, as does the dual magnetization in Eq.~(\ref{supp:eq:m2decay}), and the finite-size gap closes as $\Delta E\simeq\pi/(2L)$. This is the $1/L$ splitting of two decoupled Ising CFTs, whose first excitation is the $\sigma\otimes\sigma$ primary of total scaling dimension $\tfrac18+\tfrac18$, rather than the exponentially small splitting of a symmetry-broken doublet. The Hellmann--Feynman argument below relates this distinction to whether the self-dual transition is continuous or first order.

\subsubsection{Hellmann--Feynman criterion for duality breaking}
The relation between the duality order parameter and the order of the transition follows from Hellmann--Feynman and holds in any dimension. For the interpolation $H(s)=(1-s)H_{\rm SPT}+s\,H_{\rm para}$,
\begin{equation}\label{supp:eq:HF}
\mathcal M\coloneqq\frac{\partial H}{\partial s}=H_{\rm para}-H_{\rm SPT}=\sum_i \mathcal M_i.
\end{equation}
The $U$-odd order parameter is the field conjugate to $s$, so Hellmann--Feynman gives $\langle\mathcal M\rangle=dE_0/ds$. Here $E_0(s)$ is the \emph{total} ground-state energy, while $e_0(s)\coloneqq E_0(s)/N$ is its density per site for $N$ qubits; translation invariance gives $\langle \mathcal M_i\rangle=e_0'(s)$. Because $E_0(s)$ is the minimum of functions linear in $s$, it is concave and has one-sided derivatives everywhere. Unitarity of $U$ also implies the exact reflection symmetry $E_0(s)=E_0(1-s)$. At $s=\tfrac12$, the slope must therefore either vanish continuously, $e_0'(\tfrac12)=0$, or jump antisymmetrically across a concave kink, $e_0'(\tfrac12^{\pm})=\mp|e_0'|$. In the latter case the transition is first order, and the two coexisting ground states are exchanged by $U$ and carry opposite order-parameter densities,
\begin{equation}\label{supp:eq:mU-latent}
\langle \mathcal M_i\rangle_{\pm}=\mp\,m_U,\qquad m_U\coloneqq\big|e_0'(\tfrac12^-)\big| .
\end{equation}
Thus the latent slope discontinuity is the $U$ order parameter.

The same argument applies at every temperature, with the free energy $F(s,T)\coloneqq-T\ln\mathrm{Tr}\,e^{-\beta H(s)}$ replacing $E_0$. Its finite-temperature Hellmann--Feynman identity is
\begin{equation}\label{supp:eq:HF-T}
\frac{\partial F}{\partial s}\bigg|_{T}
=\big\langle \partial_sH(s)\big\rangle_{\beta}
\coloneqq\frac{\mathrm{Tr}\big[\partial_sH(s)\,e^{-\beta H(s)}\big]}{\mathrm{Tr}\,e^{-\beta H(s)}}.
\end{equation}
Duhamel's formula proves Eq.~(\ref{supp:eq:HF-T}) without requiring $[H,\partial_sH]=0$: $\partial_s e^{-\beta H}=-\int_0^\beta d\tau\,e^{-(\beta-\tau)H}\,\partial_sH\,e^{-\tau H}$. Cyclicity makes the traced integrand independent of $\tau$, and hence $\partial_s\,\mathrm{Tr}\,e^{-\beta H}=-\beta\,\mathrm{Tr}\big[(\partial_sH)\,e^{-\beta H}\big]$ exactly. The quantity differentiated is the \emph{free} energy: $\langle\partial_sH\rangle_\beta$ is the generalized force conjugate to $s$, the \emph{adiabatic} derivative $(\partial\langle H\rangle/\partial s)_S$, not the isothermal derivative of the energy.

The conjugate operator is also exactly duality odd. Differentiating $UH(s)U^\dagger=H(1-s)$ gives $U\,\partial_sH(s)\,U^\dagger=-(\partial_sH)(1-s)$, and hence $U\,\partial_sH(\tfrac12)\,U^\dagger=-\partial_sH(\tfrac12)$. Equation~(\ref{supp:eq:HF}) expresses $\partial_sH=\mathcal M=\sum_i\mathcal M_i$ as a sum of the local $U$-odd densities in Eq.~(\ref{supp:eq:Ol}). Thus $s-\tfrac12$ acts as a longitudinal field for the duality symmetry, just as a magnetic field does for a ferromagnetic $\mathbb Z_2$ symmetry. The free energy is concave in $s$ [$\ln\mathrm{Tr}\,e^{A+sB}$ is convex in a linear coupling] and is exactly symmetric under $s\leftrightarrow1-s$, so its one-sided derivatives exist and are antisymmetric about $s=\tfrac12$.

The $U$-symmetric Gibbs state has $\langle\mathcal M\rangle_\beta=0$ exactly at $s=\tfrac12$; spontaneous symmetry breaking is instead diagnosed by the two-sided limits. A broken phase has
\begin{equation}\label{supp:eq:mU-T}
\lim_{s\to\frac12^\pm}\frac{\partial f}{\partial s}\bigg|_{T}=\mp\,m_U(T)\neq0,
\qquad
\langle \mathcal M_i\rangle_{\pm}=\mp\,m_U(T)
\end{equation}
where $f\coloneqq F/N$. The discontinuity of the first derivative is the thermodynamic definition of a first-order transition, and the two coexisting Gibbs branches are exchanged by $U$ and carry opposite $U$-odd densities. If instead $m_U(T)=0$, then $F(\cdot,T)$ is $C^1$ at the self-dual point, $U$ is unbroken, and a transition crossing the self-dual line there is continuous. For the two $U$-conjugate branches considered here, spontaneous duality breaking, a nonzero conjugate order parameter, and a first-order wall pinned at $s=\tfrac12$ are equivalent, provided that the chosen $U$-odd density overlaps the broken order. Equation~(\ref{supp:eq:mU-T}) reduces to Eq.~(\ref{supp:eq:mU-latent}) at $T=0$. At finite temperature it identifies the first-order segment of Fig.~3(b) of the Letter, and the cluster-side ordered--ordered coexistence below $T_x$ in Sec.~\ref{supp:sm:wedge}, as regions with $m_U(T)>0$.

The dual interpolation obeys the corresponding statement, with one refinement due to non-invertibility. Its conjugate operator is $\tilde{\mathcal M}\coloneqq\partial_s\tilde H=\sum_i\tilde{\mathcal M}_i$, where $\tilde{\mathcal M}_i\coloneqq B_i-X_i$ and $B_i$ denotes the dual-side stabilizer: $Z_{i-1}Z_{i+1}$ on the chain and the plaquette flux $B_p$ in the gauge theory. The KT map intertwines the conjugate operators term by term,
\(\tilde D\,(X_i-X_iB_i)=(X_i-B_i)\,\tilde D\),
as used again in Sec.~\ref{supp:sm:3d-selfdual}. Kramers--Wannier duality on each sublattice chain, or Wegner duality in the gauge theory, makes $\partial_s\tilde H$ odd under the dual model's self-duality.

On the unprojected Hilbert space this self-duality is non-invertible. It becomes an exact symmetry in the all-plus full-fusion sector selected by $P_+$, where
\(F_+(s,T)\coloneqq-T\ln\Tr[P_+e^{-\beta\tilde H(s)}]\)
is symmetric under $s\leftrightarrow1-s$. Equation~(\ref{supp:eq:HF-T}) then holds with $P_+$ inserted and gives $\partial_sF_+=\langle\partial_s\tilde H\rangle_+$. The dual wall is therefore first order exactly when the two-sided jump of
\(\langle\tilde{\mathcal M}\rangle/N_l=\langle B_p\rangle-\langle X_l\rangle\)
is nonzero, with $\tilde{\mathcal M}$ restricted to the measured gauge copy and $N_l=3L^3$. The jump directly measured in Sec.~\ref{supp:sm:arm-qmc} is the latent slope of the \emph{full} free energy, whose first-order wall lies slightly off the self-dual line ($s_c>\tfrac12$); it is not the projected quantity. The projected free energy $F_+$ is exactly self-dual, so its duality-breaking wall---if present---is pinned at $s=\tfrac12$. The two must not be identified.

When all charge sectors are included, they break the exact $s\leftrightarrow1-s$ symmetry. At sufficiently low temperature their gapped-defect contribution shifts the crossing by the asymptotic amount derived in Sec.~\ref{supp:sm:3d-wall}, without removing the transverse first-order crossing. Equation~(\ref{supp:eq:HF-T}) still identifies the jump of $\langle\tilde{\mathcal M}\rangle/N_l$ with the kink of the full free energy.

\begin{figure}[t]
\centering
\includegraphics[width=0.9\columnwidth]{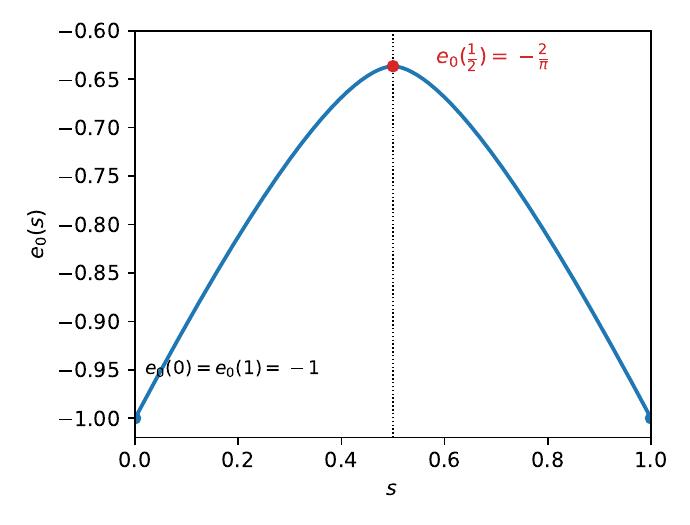}
\caption{Exact ground-state energy density of the one-dimensional interpolation, $e_0(s)=-\frac{2}{\pi}E(4s(1-s))$ [Eq.~(\ref{supp:eq:e0-1d})], manifestly symmetric under $s\leftrightarrow1-s$, with $e_0(0)=e_0(1)=-1$ and $e_0(\tfrac12)=-2/\pi$.}
\label{supp:fig:e0-1d}
\end{figure}

In one dimension the dichotomy can be settled in closed form. Summing the vacuum energies of the two chains of Sec.~\ref{supp:sm:ff}, $E_0(s)=-\sum_{k\in{\rm NS}}\varepsilon_k(s)$, and rewriting the dispersion~(\ref{supp:eq:bogoliubov}) as $\varepsilon_k=2\sqrt{1-\mathfrak m\cos^2(k/2)}$ with $\mathfrak m\coloneqq4s(1-s)$, the thermodynamic-limit density per qubit is a complete elliptic integral of the second kind,
\begin{equation}\label{supp:eq:e0-1d}
e_0(s)=-\int_0^{2\pi}\!\frac{dk}{2\pi}\,\sqrt{1-\mathfrak m\cos^2\tfrac k2}
=-\frac{2}{\pi}\,E(\mathfrak m),
\end{equation}
manifestly symmetric under $s\leftrightarrow1-s$ (through $\mathfrak m$). Its derivative is exactly
\begin{equation}\label{supp:eq:e0p-1d}
\begin{aligned}
e_0'(s)&=-\frac{4(1-2s)}{\pi\mathfrak m}\big[E(\mathfrak m)-K(\mathfrak m)\big]\\
&\xrightarrow[\;s\to\tfrac12\;]{}\;
\frac{4\delta}{\pi}\ln\frac{4}{e\,|\delta|}\;\to\;0,
\end{aligned}
\end{equation}
with $\delta\coloneqq1-2s$ and $K$ the complete elliptic integral of the first kind, for which $K(\mathfrak m)\simeq\ln(4/|\delta|)$ as $\mathfrak m\to1$. The absolute value keeps the logarithm real on both sides, while the factor $\delta$ makes the asymptotic odd, as required by $E_0(s)=E_0(1-s)$. The slope vanishes continuously as $\delta\ln(1/\delta)$, with only a logarithmically divergent curvature. Thus the continuous two-Ising transition has no latent kink: $\langle \mathcal M_i\rangle=e_0'(s)\to0$, and $U$ is unbroken, consistently with Eq.~(\ref{supp:eq:chi2}). Equations~(\ref{supp:eq:e0-1d}) and~(\ref{supp:eq:e0p-1d}) reproduce exact diagonalization and the finite-$L$ momentum sums to machine precision. Away from $s=\tfrac12$, Eq.~(\ref{supp:eq:e0p-1d}) also gives the exact profile $\langle \mathcal M_i\rangle=e_0'(s)$ of the $U$-odd density. Figure~\ref{supp:fig:e0-1d} plots the resulting smooth, concave, and symmetric energy density.

The three-dimensional interpolation instead has a first-order self-dual point, as shown in Sec.~\ref{supp:sm:3d-selfdual}.

\section{Three dimensions: derivations and numerics}
\label{supp:sm:3d}

\subsection{Geometry of the one-form flux twist}
\label{supp:sm:3d-twist-geom}
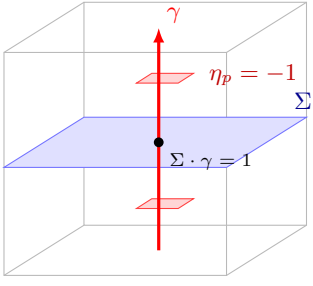
\begin{figure}[t]
\centering
\begin{tikzpicture}[scale=0.92,>={Latex[length=2mm]}]
  \def\dx{1.15}\def\dy{0.72}
  \coordinate (FBL) at (0,0); \coordinate (FBR) at (3.2,0);
  \coordinate (FTR) at (3.2,3.2); \coordinate (FTL) at (0,3.2);
  \coordinate (BBL) at (\dx,\dy); \coordinate (BBR) at (3.2+\dx,\dy);
  \coordinate (BTR) at (3.2+\dx,3.2+\dy); \coordinate (BTL) at (\dx,3.2+\dy);
  \draw[gray!55] (FBL)--(FBR)--(FTR)--(FTL)--cycle;
  \draw[gray!55] (BBL)--(BBR)--(BTR)--(BTL)--cycle;
  \draw[gray!55] (FBL)--(BBL) (FBR)--(BBR) (FTR)--(BTR) (FTL)--(BTL);
  \fill[blue!12] (0,1.55)--(3.2,1.55)--(3.2+\dx,1.55+\dy)--(\dx,1.55+\dy)--cycle;
  \draw[blue!55] (0,1.55)--(3.2,1.55)--(3.2+\dx,1.55+\dy)--(\dx,1.55+\dy)--cycle;
  \node[blue!55!black] at (3.2+\dx-0.05,1.55+\dy+0.24) {$\Sigma$};
  \pgfmathsetmacro{\gsx}{1.65+0.5*\dx}
  \pgfmathsetmacro{\gby}{0.5*\dy}
  \foreach \zz in {0.7,2.5} {
    \pgfmathsetmacro{\zs}{\zz+0.5*\dy}
    \fill[red!16] (\gsx-0.32,\zs-0.10)--(\gsx+0.28,\zs-0.10)--(\gsx+0.28+0.22,\zs+0.04)--(\gsx-0.32+0.22,\zs+0.04)--cycle;
    \draw[red!75] (\gsx-0.32,\zs-0.10)--(\gsx+0.28,\zs-0.10)--(\gsx+0.28+0.22,\zs+0.04)--(\gsx-0.32+0.22,\zs+0.04)--cycle;
  }
  \node[red!75!black,right] at (\gsx+0.6,2.5+0.5*\dy+0.02) {$\eta_p=-1$};
  \draw[red,very thick,->] (\gsx,\gby) -- (\gsx,3.2+\gby);
  \node[red] at (\gsx+0.22,3.2+\gby+0.18) {$\gamma$};
  \pgfmathsetmacro{\csy}{1.55+0.5*\dy}
  \fill[black] (\gsx,\csy) circle (2pt);
  \node[below right] at (\gsx+0.03,\csy-0.03) {\scriptsize $\Sigma\cdot\gamma=1$};
\end{tikzpicture}
\caption{Geometry of the one-form flux twist on the $3$-torus. The flux loop $\gamma$ (red, a noncontractible $1$-cycle) threads the measured membrane $\Sigma$ (blue, a noncontractible $2$-cycle of plaquettes) once, $\Sigma\cdot\gamma=1$ (dot). The twist is the flat, non-exact two-cocycle $\eta_p=-1$ on the stack of plaquettes pierced by $\gamma$ (two shown), Poincar\'e-dual to $[\gamma]$---\emph{not} a coboundary of link sign flips. At $T=0$ the twisted membrane collects the braiding phase $\langle S_\Sigma^{(1)}\rangle_{\rm tw}=(-1)^{\Sigma\cdot\gamma}$.}
\label{supp:fig:twist-geom}
\end{figure}

The twist that diagnoses the three-dimensional SPT order is the flat $\mathbb Z_2$ two-cocycle Poincar\'e-dual to a noncontractible flux loop $\gamma$ ($\eta_p\coloneqq-1$ on the plaquettes pierced by $\gamma$), inserted as $X_pB_p\to\eta_pX_pB_p$; the membrane $S_\Sigma^{(1)}\coloneqq\prod_{p\in\Sigma}X_p$ then acquires the braiding phase $(-1)^{\Sigma\cdot\gamma}$ (Fig.~\ref{supp:fig:twist-geom}).

\subsection{Shape of the deconfinement boundary: exact statements}
\label{supp:sm:3d-boundary}

The link-qubit sector of the dual model, $-(1-s)\sum_p B_p-s\sum_l X_l$ with $B_p=\prod_{l\in\partial p}Z_l$ its plaquette flux operator, is the three-dimensional $\mathbb Z_2$ Ising gauge theory in a transverse field of strength $h^x\coloneqq s/(1-s)$; its deconfinement boundary $T_c(s)$, introduced in the Letter, is a property of the plain gauge theory alone---no strong-symmetry construction enters. At zero field ($s=0$) Wegner duality~\cite{Wegner} fixes it to the three-dimensional Ising transition, $e^{-2/T_c(0)}=\tanh K_c^{\rm Ising}$ with $K_c^{\rm Ising}\approx0.2216$, giving
\begin{equation}\label{supp:eq:Tc}
T_c(0)\approx1.3133,
\end{equation}
and the boundary terminates at the $T=0$ first-order transition, pinned by self-duality exactly at $h^x_c=1$ (assuming a unique transition), i.e.\ $s=\tfrac12$~\cite{ReissSchmidt}.

\subsubsection{Endpoints, evenness, and the exact initial slope}
\label{supp:sm:3d-exact}
The endpoints of $T_c(s)$ are Eq.~(\ref{supp:eq:Tc}) at $s=0$ and the self-dual point $s=\tfrac12$ at $T=0$. Its initial slope also follows exactly. The conjugation $C\coloneqq\prod_l Z_l$ obeys $CB_pC=B_p$ and $CX_lC=-X_l$, so $C\,\tilde H(s)\,C$ reverses the field and makes the spectrum even in $h^x$. Writing $\tilde H(s)=(1-s)\big[-\sum_p B_p-h^x\sum_l X_l\big]$ rescales the coupling to unity, giving
\begin{equation}\label{supp:eq:rescale}
T_c(s)=(1-s)\,\tilde T_c(h^x),\qquad h^x=\tfrac{s}{1-s},
\end{equation}
with $\tilde T_c$ the critical temperature of the unit-coupling model and an even function of $h^x$ by the conjugation. Hence $\tilde T_c{}'(0)=0$ and
\begin{equation}\label{supp:eq:slope}
\frac{dT_c}{ds}\Big|_{s=0}=-\tilde T_c(0)=-T_c(0).
\end{equation}
The initial descent is entirely the trivial rescaling $(1-s)$, with genuine field corrections appearing only at $O(s^2)$, $T_c(s)=T_c(0)\,[\,1-s-a\,s^2+O(s^3)\,]$, a single constant $a$ controlling the leading curvature---left undetermined by the evenness, which forbids only the linear term (it does \emph{not} fix the sign of $a$), and physically expected positive as the field lowers $T_c$. In Sec.~\ref{supp:sm:arm-qmc} we determine it numerically, $a=0.21(2)$.

\subsubsection{The two arms and the persistence of first order}
\label{supp:sm:3d-arms}
The order of the transition differs between the two arms because the effective dimension changes. Wherever the thermal deconfinement boundary is continuous, the finite imaginary-time direction leaves the universality class of the three-dimensional classical $\mathbb Z_2$ gauge theory, or equivalently the three-dimensional Ising class. This does not preclude a separate finite-temperature first-order line. At the endpoint $(s,T)=(\tfrac12,0)$, the imaginary-time direction is infinite and the model becomes the self-dual four-dimensional $\mathbb Z_2$ gauge theory, whose transition is weakly first order~\cite{ReissSchmidt,BalianDrouffeItzykson,CreutzJacobsRebbi}. Both coexisting branches there have finite correlation length. At sufficiently large $\beta$, their free energies receive only smooth, exponentially small corrections, which cannot remove a transverse crossing of two analytic branches with unequal $\partial_s f$. A first-order line must therefore emanate from $(\tfrac12,0)$.

\subsubsection{Asymptotic pinning of the wall and the scale of $T^*$}
\label{supp:sm:3d-wall}
Wegner self-duality acts exactly on the all-plus full-fusion sector selected by $P_+$ and exchanges the deconfined and confined branches. Writing their projected free-energy densities as $f_{{\rm dec},+}$ and $f_{{\rm conf},+}$,
\begin{equation}
f_{{\rm dec},+}(s,T)=f_{{\rm conf},+}(1-s,T),
\end{equation}
so, whenever both branches exist, they coexist at $s=\tfrac12$. The full trace also sums the extensive charge sectors and need not be self-dual: at $s=1$ the model $-\sum_lX_l$ is analytic at every $T$, whereas the $s=0$ dome is singular.

To quantify the low-temperature displacement after restoring all charge sectors, let $a={\rm dec},{\rm conf}$ label the two branches and define the charge-sector correction by
\begin{equation}
\delta f_a(s,T)\coloneqq f_a(s,T)-f_{a,+}(s,T).
\end{equation}
Both branches at $(\tfrac12,0)$ are gapped. If $\Delta_a>0$ is the lowest allowed charge-defect excitation energy in branch $a$, the dilute-defect expansion gives
\begin{equation}
\delta f_a(\tfrac12,T)=O\!\left(T e^{-\Delta_a/T}\right).
\end{equation}
Define the nonzero latent slope
\begin{equation}
\ell(T)\coloneqq\left.\partial_s\!\left(f_{{\rm dec},+}-f_{{\rm conf},+}\right)\right|_{s=1/2}.
\end{equation}
Expanding the resulting coexistence condition about $s=\tfrac12$ gives
\begin{align}
0={}&\ell(T)\bigl[s_c(T)-\tfrac12\bigr]
+\delta f_{\rm dec}(\tfrac12,T)-\delta f_{\rm conf}(\tfrac12,T)\nonumber\\
&+O\!\left([s_c-\tfrac12]^2\right),
\end{align}
and hence
\begin{align}
s_c(T)-\tfrac12
={}&-\frac{\delta f_{\rm dec}(\tfrac12,T)
              -\delta f_{\rm conf}(\tfrac12,T)}
             {\ell(T)}
 +o(\delta f)\nonumber\\
={}&O\!\left(T e^{-\Delta_q/T}\right),
\qquad \Delta_q\coloneqq\min_a\Delta_a>0 .
\end{align}
This result is asymptotic as $T\to0$. It fixes neither the prefactor nor the onset temperature and makes no quantitative prediction over the accessible range $0.04\le T\le0.22$. Nor does it pin the tricritical point to the self-dual line: the segment ends at a generally displaced point $(s^*,T^*)$, determined below.

The SPT-side interpolation $H(s)$ has a related structure pinned \emph{exactly} at $s=\tfrac12$ by the unitary $U$. The duality symmetry is spontaneously broken across this wall, which opens into the self-dual frozen wedge studied in Sec.~\ref{supp:sm:wedge}.

\subsection{The self-dual point in three dimensions: a broken duality symmetry}
\label{supp:sm:3d-selfdual}

At $s=\tfrac12$ the entangler is again a genuine $\mathbb Z_2$ symmetry,
$[U,H(\tfrac12)]=0$. The Hellmann--Feynman result of Sec.~\ref{supp:sm:selfdual}
then implies that $U$ breaks spontaneously if and only if the ground-state energy
density has a kink at the self-dual point, with order parameter
$m_U=|e_0'(\tfrac12^-)|$ [Eq.~(\ref{supp:eq:mU-latent})]. The kink absent in one
dimension is present in three dimensions.

Within the all-plus full-fusion sector selected by $P_+$, the KT isometry makes the RBH interpolation $H(s)$ isospectral to two decoupled transverse-field $\mathbb Z_2$ gauge theories, and the ground state on each branch can be chosen in that sector. The contractible star operators $S^{(1)}_v=\prod_{l\ni v}X_l$ commute with $H(s)$ because every plaquette shares either zero or two links with a star. They are therefore superselection labels, and the ground state has a definite charge configuration.

Perron--Frobenius establishes that this ground state is \emph{contractible-charge-free} for every $s>0$, without assuming the absence of a level crossing. In the $Z$ basis, the off-diagonal single-spin-flip matrix elements are $-[s+(1-s)B_i(z)]$. For $s>\tfrac12$ they are strictly negative for both eigenvalues $B_i(z)=\pm1$; at $s=\tfrac12$, the amplitude for $B_i(z)=-1$ is $-(2s-1)=0$. The flip graph is irreducible for $s>\tfrac12$, so its ground state is unique and has strictly positive amplitudes. Every commuting $X$-type symmetry, including the contractible stars and the noncontractible membrane generators, permutes the $Z$-basis configurations. It therefore maps the positive ground state to another positive ground state, which uniqueness forces to be the same state. Hence the ground state lies in the all-plus $P_+$ sector. The branch at $s<\tfrac12$ follows from $U^\dagger H(s)U=H(1-s)$ and $US_aU^\dagger=S_a$. One-sided limits establish all-plus ground-state branches at $s=\tfrac12$; we do not claim uniqueness exactly there, where the first-order transition makes the two $U$-conjugate branches degenerate. In the electric picture, the contractible-charge result follows from the same energetics: a charge $S^{(1)}_v=-1$ terminates an open string of $x_l=-1$ links, and $-s\sum_lX_l$ gives the string a tension proportional to $s$, positive for $0<s<1$ and vanishing only as $s\to0$.

On this all-plus branch, the cluster-model ground-state energy density equals that of the transverse-field gauge theory. Its Euclidean path integral is the four-dimensional classical $\mathbb Z_2$ gauge theory~\cite{FradkinSusskind}, which is self-dual at $s=\tfrac12$~\cite{Wegner,BalianDrouffeItzyksonII}. Gauge-invariant Monte Carlo finds hysteresis and two-phase coexistence there~\cite{CreutzJacobsRebbi}, while strong-coupling and mean-field analyses give the same weakly first-order transition~\cite{BalianDrouffeItzykson}. Its weakness is controlled by the large four-dimensional correlation length $\xi_4$~\cite{ReissSchmidt}. Equation~(\ref{supp:eq:mU-latent}) consequently gives $m_U=|e_0'(\tfrac12^-)|\neq0$. Thus $U$ is weakly but spontaneously broken at the three-dimensional self-dual point, and the diagnostic in Eq.~(\ref{supp:eq:m2U}) saturates at $m_U^2>0$ rather than decaying as $2/L$.

Because $U$ is a genuine unitary on the unprojected Hilbert space, the cluster-model free energy obeys $F(s,T)=F(1-s,T)$ at every temperature. This differs from the gauge theory of Sec.~\ref{supp:sm:3d-wall}, where Wegner self-duality gives an exact finite-size identity only in the all-plus $P_+$ sector. At the first-order point, the two gapped branches receive only smooth, exponentially small thermal corrections. The transverse crossing therefore persists under weak heating, as in Sec.~\ref{supp:sm:3d-arms}, and the symmetry of $F$ pins it exactly to $s=\tfrac12$. On the gauge side the crossing approaches that value only asymptotically. Wherever the two $U$-conjugate phases remain the competing branches, the RBH phase diagram thus contains a first-order segment across which the local $U$-odd density $\langle\mathcal M_i\rangle$ jumps by $2m_U(T)$.

This two-branch argument permits a third branch to intervene at $T>0$. Section~\ref{supp:sm:wedge} shows that such a branch does occur: direct ordered--ordered coexistence survives only below $T_x\lesssim0.01$---a heuristic estimate from an $L=3,4$, simplified entropy--energy balance [Eq.~(\ref{supp:eq:Tx-sm})], not a controlled bound---above which the wall opens into a self-dual frozen wedge. Equations~(\ref{supp:eq:HF-T})--(\ref{supp:eq:mU-T}) identify the jump with the discontinuity of the isothermal derivative $\partial_sf$. The wall is therefore first order in the thermodynamic sense of a free-energy kink, and spontaneous breaking of $U$ at temperature $T$ is equivalent to ordered--ordered coexistence extending to that temperature.

The KT intertwining relation $\tilde D\,\mathcal M=\tilde{\mathcal M}\,\tilde D$ maps the two latent-slope operators within the all-plus $P_+$ sector. It does not identify the corresponding first-order segments of the two models, whose locations may differ because their charge sectors differ. Moreover, this wall is not a revival of finite-temperature SPT order. The broken symmetry is the ordinary, zero-form, non-onsite duality $\mathbb Z_2$ present only on the self-dual line. The one-form twisted membrane shows no finite-temperature revival at the fixed points, in the rigorous high-temperature regime, or in the accessible lower-temperature numerical window.

\subsection{QMC methods and the SPT-side phase diagram}
\label{supp:sm:tstar-qmc}

We determine the finite-temperature phase boundaries using two QMC implementations. On the gauge side we use the continuous-time algorithm of Wu, Deng, and Prokof'ev~\cite{WuDengProkofev}, as implemented in ParaToric~\cite{ParaToric}, in the $Z$ basis for the unit-coupling Hamiltonian $-J\sum_pB_p-h^x\sum_lX_l$ with $J=1$. In the parameter range used below, we validated it against a classical Metropolis benchmark and an independent permutation-matrix-representation QMC~\cite{PMRQMC}; all observables agree within $1.7\sigma$ at an interacting point. Temperatures convert to the convention of the Letter as $T=(1-s)\tilde T$, and hence $T=\tilde T/2$ at $s=\tfrac12$. One \emph{sweep} denotes one attempted local update per space-time degree of freedom.

The cluster-side wedge calculations in Sec.~\ref{supp:sm:wedge} instead use discrete-time worldline QMC with imaginary-time step $\Delta\tau$ and $N_t\coloneqq\beta/\Delta\tau$ slices. Observables are extrapolated to $\Delta\tau\to0$, or evaluated at sufficiently small $\Delta\tau$ that the residual Trotter bias is below the statistical error; the gauge-side continuous-time calculation has no Trotter error. Parenthetic uncertainties in the boundary tables are $1\sigma$ values obtained by adding the statistical error, crossing or fit spread, and finite-size extrapolation in quadrature.

\subsubsection{The twisted membrane across the $(s,T)$ plane}
The twisted membrane obeys an exact area law at the fixed points,
\(\langle S_\Sigma^{(1)}\rangle_{\rm tw}=\mp\tanh^{|\Sigma|}\!\beta\)
with $|\Sigma|\sim L^2$. The same suppression is rigorous in the convergent high-temperature regime derived below. In the remaining accessible finite-size window, permutation-matrix-representation QMC of the cluster interpolation~\cite{PMRQMC} finds no finite-temperature revival.

The interpolation Gibbs weight is stoquastic, and hence sign-free, for $s\ge\tfrac12$; an exact twisted reflection determines the $s<\tfrac12$ side. Write
\(H_{\rm tw}(s)\coloneqq-(1-s)(\sum_lK_l+\sum_p\eta_pK_p)-s\sum_iX_i\),
with $K_i\coloneqq X_iB_i$. Conjugation by $U$ alone moves the cocycle $\eta_p$ from the cluster terms to the paramagnetic fields. The diagonal sign unitary
\(W_\eta\coloneqq\prod_{p:\,\eta_p=-1}Z_p\)
restores the original placement because $W_\eta K_pW_\eta^\dagger=\eta_pK_p$ and $W_\eta X_pW_\eta^\dagger=\eta_pX_p$, while it leaves the link terms fixed. Therefore
\((W_\eta U)H_{\rm tw}(s)(W_\eta U)^\dagger=H_{\rm tw}(1-s)\)
and
\((W_\eta U)S_\Sigma^{(1)}(W_\eta U)^\dagger=(-1)^{\Sigma\cdot\gamma}S_\Sigma^{(1)}\).
For odd $\Sigma\cdot\gamma$ this gives the exact finite-size, finite-temperature identity
\(\langle S_\Sigma^{(1)}\rangle_{\rm tw}(s)=-\langle S_\Sigma^{(1)}\rangle_{\rm tw}(1-s)\),
which vanishes at $s=\tfrac12$. We therefore simulate $s\ge\tfrac12$ and infer the other half from this relation.

\begin{figure}[t]
\centering
\includegraphics[width=0.8\columnwidth]{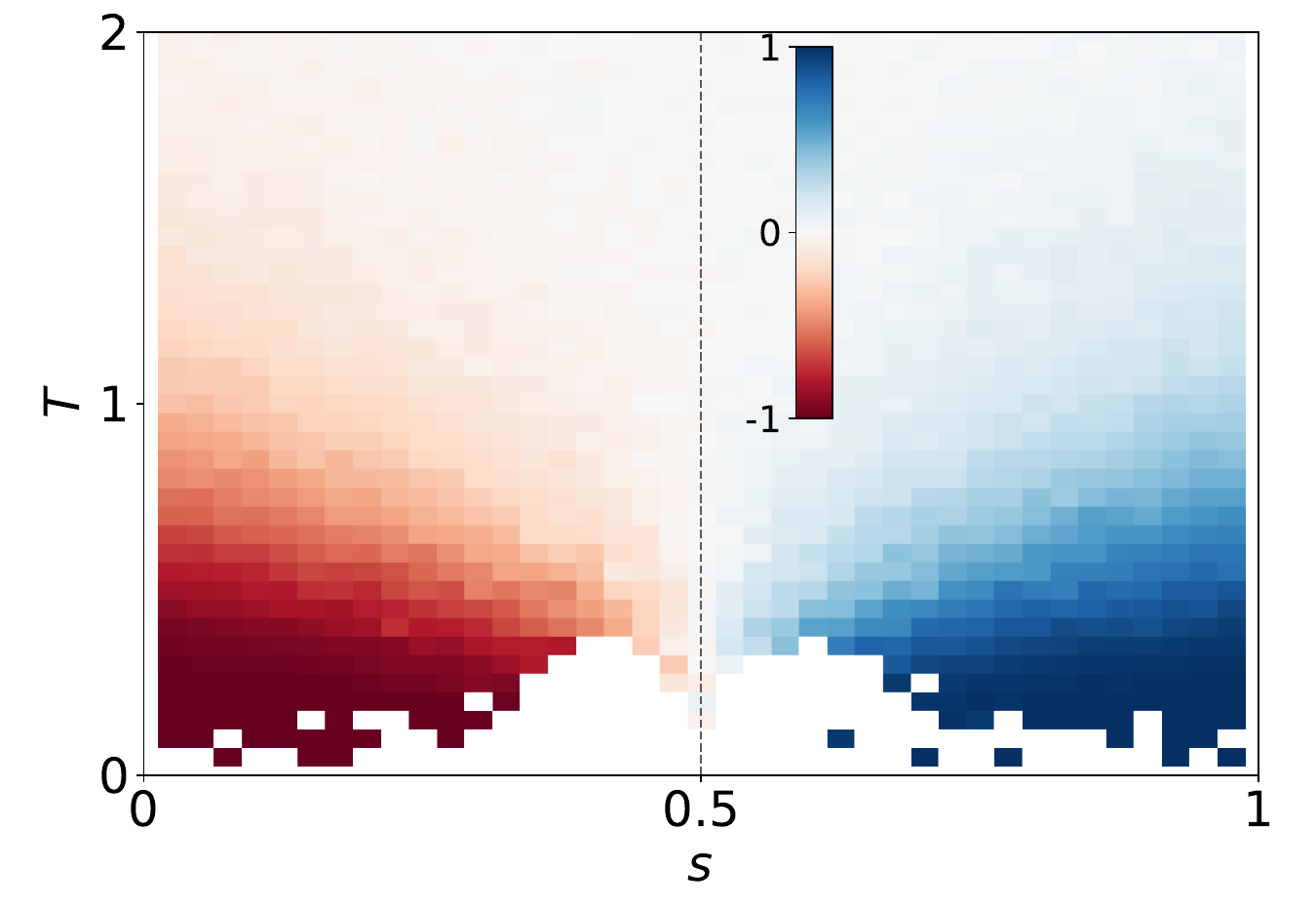}
\caption{One-form flux-twisted membrane $\langle S_\Sigma^{(1)}\rangle_{\rm tw}(s,T)$ of the cluster interpolation $H(s)$ (PMR-QMC~\cite{PMRQMC}, $L=2$, $N=6L^3=48$ qubits, $64$ independent chains). As $T\to0$ the endpoints saturate the exact plateaus $\mp1$: red ($-1$) across the SPT phase, blue ($+1$) across the trivial phase, with the jump at $s=\tfrac12$ (dashed). White marks the region where the measurement-basis sign problem prevents a reliable estimate; as $T\to0$ the reliable region narrows to the fixed-point columns, precisely where the intermediate-$s$ order is itself area-law suppressed.}
\label{supp:fig:mtw-sm}
\end{figure}

The measurement rotates the membrane to the diagonal operator $\prod_{p\in\Sigma}Z_p$. In three dimensions this rotation makes the $2L^2$ link stabilizers touching $\Sigma$ off diagonal, creating a sign problem that grows with the membrane \emph{area}. Reliable estimates are consequently restricted to $T\gtrsim T_\ast(L)$. The fixed-point columns $s=0,1$ are \emph{exactly} sign-free at every temperature, since the rotation creates no off-diagonal link terms there; their neighborhoods stay numerically accessible only over a finite low-temperature range that widens toward the columns. Figure~\ref{supp:fig:mtw-sm} shows the result for $L=2$. As $T\to0$, the quantized step $\mathrm{sgn}(s-\tfrac12)$ emerges and the endpoints approach $\mp1$. For $L=2,3,4$, QMC reproduces the exact endpoint law $\mp\tanh^{L^2}\!\beta$ within statistical error, while the apparently ordered region contracts increasingly rapidly toward the $T=0$ axis. These data underlie the red and blue $T=0$ lines in Fig.~3(a) of the Letter and show that the bare, weak-Gibbs one-form SPT order does not survive at finite temperature in the accessible window.

We now establish the high-temperature bound. At small $\beta$ the suppression is rigorous, uniformly in $L$, $s$, and the twist. Expand both traces of $\langle S_\Sigma^{(1)}\rangle_{\rm tw}\coloneqq\mathrm{Tr}[S_\Sigma^{(1)}e^{-\beta H_{\rm tw}}]/\mathrm{Tr}[e^{-\beta H_{\rm tw}}]$ in the high-temperature (Pauli) series in the local terms of $H_{\rm tw}$. A numerator term is a product of local Pauli factors whose trace is nonzero only if the total operator is proportional to the identity; since $S_\Sigma^{(1)}=\prod_{p\in\Sigma}X_p$ carries an $X$ on each of the $|\Sigma|$ plaquettes of $\Sigma$, and the only terms of $H_{\rm tw}$ supported by $X_p$ are $-sX_p$ and $-(1-s)\eta_pX_pB_p$ (combined transverse weight $s+(1-s)=1$), every $p\in\Sigma$ must receive at least one insertion. For $C\beta<1$---$C$ a lattice constant fixed by the coordination number, independent of $L$---the Kotecky--Preiss criterion~\cite{KoteckyPreiss} renders the cluster expansions of numerator and denominator absolutely convergent; the extensive vacuum clusters decorate both traces identically---each contributing a volume factor $e^{O(N)}$---and cancel in the ratio (whose denominator is in any case positive, $\mathrm{Tr}\,e^{-\beta H_{\rm tw}}\ge2^N$ since $\mathrm{Tr}\,H_{\rm tw}=0$), leaving a convergent sum over clusters pinned to $\Sigma$. As each of its $|\Sigma|$ plaquettes must be activated, this gives
\begin{equation}\label{supp:eq:hiT-membrane}
\big|\langle S_\Sigma^{(1)}\rangle_{\rm tw}\big|\le(C\beta)^{|\Sigma|}=e^{-a(\beta)|\Sigma|},\qquad a(\beta)\coloneqq\ln\tfrac{1}{C\beta}>0,
\end{equation}
uniformly in $L$ and $s$. This rigorous high-temperature suppression, together with the exact all-temperature fixed-point law, brackets the bare membrane; only the low-temperature interior---the white region of Fig.~\ref{supp:fig:mtw-sm}, where the measurement-basis sign problem precludes a reliable estimate---rests on the finite-size QMC alone, which shows no revival across the accessible window.

\subsection{The self-dual frozen wedge}
\label{supp:sm:wedge}

\subsubsection{The frozen manifold}
On the SPT side, the two competing \emph{link} families of $H(s)$ commute:
$[X_lB_l,X_{l'}]=0$ for all $l,l'$, because the dressing $B_l$ acts on plaquette qubits whereas $X_{l'}$ acts on link qubits. By contrast, the plaquette families $\{X_pB_p\}$ and $\{X_p\}$ do not commute with both link families and enter below through virtual dressing. The commuting stabilizer family $\{X_lB_l\}\cup\{X_l\}$ has simultaneous $+1$ eigenstates satisfying
\begin{equation}
X_l|F\rangle=X_lB_l|F\rangle=+|F\rangle\quad\forall\,l .
\label{supp:eq:frozen-sm}
\end{equation}
These frozen configurations saturate both link families for every $s$ and have zero expectation for both plaquette families. Their classical energy $E_F\coloneqq-3L^3=-N/2$ is independent of $s$, so $\mathcal M=\partial_sH$ vanishes identically along the branch, and the entangler acts trivially, $U|F\rangle=|F\rangle$, exactly. The two commuting link families have exactly $2^{L^3+2}$ simultaneous $+1$ eigenstates: cube flips contribute $2^{L^3-1}$ configurations and the three noncontractible classes contribute a factor $2^3$. This \emph{bare configurational manifold} carries entropy
\begin{equation}
\Delta s_{\rm res}\coloneqq\frac{(L^3+2)\ln 2}{6L^3}\;\xrightarrow{\;L\to\infty\;}\;\frac{\ln2}{6}\approx0.1155
\label{supp:eq:sres}
\end{equation}
per qubit. Equation~(\ref{supp:eq:sres}) counts bare link-family eigenstates exactly, but it does not establish a degeneracy of the full Hamiltonian. The plaquette terms dress and can split these configurations. Thus Eq.~(\ref{supp:eq:sres}) describes an effective entropy only above the tunneling scale $T_x$ associated with that splitting [Eq.~(\ref{supp:eq:Tx-sm})], and higher-order cube or surface processes may quench it further as $T\to0$. Link--plaquette exchange symmetry gives a mirror variational family that saturates the two plaquette families with the same energy and count. This is likewise not an established eigenspace of the full Hamiltonian and does not change the intensive quantities below.

Virtual fluctuations strongly dress the frozen branch. At second order their contribution is $-(1-s)^2/(8s)-s^2/[8(1-s)]$ per plaquette, or $-0.0625$ per qubit at $s=\tfrac12$, and the measured branch energy $e_F\approx-0.567$ per qubit lies within $10^{-3}$ of the ordered branches. We evaluate it using sign-free worldline QMC in the mixed stabilizer basis, with link qubits in the $z$ basis and plaquette qubits in the $x$ basis. We validated the calculation against the exact decoupled limits, Trotter scaling, the mirror antisymmetry of $\mathcal M$, and, on the dual side, ParaToric~\cite{ParaToric} branch values. It gives
\begin{equation}
\Delta e\coloneqq e_F-e_{\rm ord}=+0.0015(4),\ +0.0006(3)\quad(L{=}3,4),
\label{supp:eq:de-sm}
\end{equation}
decreasing with $L$. A heuristic entropy--energy balance---which neglects the splitting within the bare manifold, its temperature-dependent effective entropy, and the entropy of the ordered branches---places the termination of the direct ordered--ordered coexistence on the self-dual line near
\begin{equation}
T_x\sim\Delta e/\Delta s_{\rm res}\lesssim 0.01 .
\label{supp:eq:Tx-sm}
\end{equation}

\subsubsection{Equilibrium at the self-dual point}
We performed $2\times10^6$-sweep runs at fixed $s=\tfrac12$, initialized in the frozen configuration and in either ordered phase. For every $T$ between $0.10$ and $0.20$ and for both $L=3$ and $L=4$, the ordered states decay onto the frozen branch. Where mixing is active, tunneling occurs in both directions; for example, one run at $T=0.12$ and $L=3$ contained $26$ ordered$\to$frozen and $25$ frozen$\to$ordered transitions. At these sizes the frozen branch is therefore the equilibrium branch on the self-dual line throughout this temperature range. The two-way tunneling is strong finite-size evidence for an intensive free-energy ordering, although it is not by itself a thermodynamic proof.

The cocycle-sector label $\Sigma_x\coloneqq\sum_{p,t}x_{p,t}/N_t$, where $x_{p,t}$ is the plaquette variable on imaginary-time slice $t$ and $N_t$ is the number of slices, shows that the residual entropy is dynamically active. Sectors remain pinned for $T\le0.08$, migrate by discrete cube flips at $T=0.10$--$0.12$, and are fully randomized for $T\ge0.20$.

\subsubsection{The wedge and its boundary}
Away from the self-dual line, the ordered free energies decrease at rate $m_{\rm ord}\coloneqq|\partial_sf_{\rm ord}|\approx0.55$, whereas the frozen branch remains flat. The same heuristic balance gives the estimated half-width
\(\delta_s(T)\coloneqq(T\Delta s_{\rm res}-\Delta e)/m_{\rm ord}\);
the wedge \(|s-\tfrac12|<\delta_s(T)\) is bounded by two ordered--frozen first-order lines. This relation is an estimate, not a controlled equality.

Equilibrium $s$ scans with $L=3,4$, $\Delta s=0.01$, and $T=0.20$--$0.40$ locate the left boundary directly; frozen and ordered initializations agree. The equilibrium profile $\langle\mathcal M\rangle(s)$ crosses the boundary with a discontinuous jump that sharpens with volume. At $T=0.20$ the step grows from $0.25$ for $L=3$ to $0.45$ for $L=4$, with boundary $s=0.465(5)$; at $T=0.25$ it grows from $0.19$ to $0.30$. For $T\ge0.30$, the profile is instead a smooth $S$ curve without volume sharpening. On the available sizes, the two first-order lines terminate at
\begin{equation}
(s,T)_{\rm end}=(\tfrac12\mp0.030(8),\;0.27(2)),
\label{supp:eq:Tend-sm}
\end{equation}
Above these endpoints, every crossing is a crossover. Along the self-dual line, the two polarizations decay smoothly from $0.83$ at $T=0.25$ to $0.37$ at $T=1.00$, with no sharp feature. Here $\langle z_l\rangle=\langle z_la_l\rangle$ are the diagonal expectations of the two competing link families in the worldline basis, and $z_l,a_l\in\{\pm1\}$ are its two commuting diagonal link variables; simultaneous saturation identifies the frozen branch. The wedge interior connects smoothly to the trivial phase. It is bounded only by the two first-order lines, like the liquid side of a liquid--gas transition, and is not an independent phase.

\subsubsection{Hysteresis and the data of Fig.~3(a) of the Letter}
Dragged $s$ chains produce three-state hysteresis loops; the inset of Fig.~3(a) of the Letter shows the $T=0.16$, $L=8$ example. Each ordered branch first decays onto the frozen plateau and only later converts into the opposite ordered phase. On the plateau, both link polarizations are large and nearly equal,
\(\langle z_l\rangle\approx\langle z_la_l\rangle\approx0.81\);
an ordered branch instead polarizes one family and disorders the other. Below $T\approx0.11$, the dragged ordered branches persist through $s=\tfrac12$ and decay onto the frozen branch only in the wings, producing the overhang of the excluded low-temperature strips. The loop changes between these forms at $T\approx0.11$--$0.13$.

The conversion and decay spinodals bracket the equilibrium boundary. In Fig.~3(a) of the Letter, the blue squares mark the strip midpoints for $L=8$ and $T=0.06$--$0.26$, together with the exact first-order point $(\tfrac12,0)$ and the endpoints in Eq.~(\ref{supp:eq:Tend-sm}). The displayed half-width is a \emph{metastability interval}, not an equilibrium statistical error bar. We exclude the $T=0.02$ and $0.04$ strips because both are dominated by the overhang, and the former also comes from an incomplete low-temperature data set.

The loops independently locate the endpoints. The strip half-width decreases from $0.007$ at $T=0.22$ to $0.001$ at $T=0.28$. Across the endpoint, the wing-maximum splitting falls as $0.20\to0.08\to0.01$ for $T=0.26\to0.28\to0.30$, so the $L=8$ hysteresis closes within the interval of Eq.~(\ref{supp:eq:Tend-sm}). The strip midpoints at $T=0.22$--$0.24$, namely $s=0.464$--$0.465$, agree with the equilibrium boundary $s=0.465(5)$. An empirical supplementary fit of the closing width to $A(T_{\rm end}-T)^\beta$ gives $T_{\rm end}=0.266(5)$ for the three-dimensional Ising exponent $\beta=0.326$ and $0.276(10)$ for $\beta=\tfrac12$. Because a dynamical hysteresis width is not an order parameter, this fit has no scaling justification and is not used to determine the endpoint; increasing metastable inflation toward low $T$ also biases the fitted value downward. Combining the loop closure with the equilibrium bracket---sharpening at $T=0.25$ and none at $T=0.30$---and extrapolating the strip midpoints gives Eq.~(\ref{supp:eq:Tend-sm}).

\subsubsection{Exclusion on the dual side}
No \emph{directly saturated} counterpart of the frozen branch exists for $\tilde H(s)$. Its competing terms share link qubits and anticommute, $\{B_p,X_l\}=0$ for $l\in\partial p$. Consequently every state at every temperature obeys
\(\langle B_p\rangle^2+\langle X_l\rangle^2\le1\),
whereas direct saturation would require both expectations to approach unity. This local bound excludes a saturated dual frozen branch, although it does not rule out every collective or entropic intermediate branch.

The dual-side loop in the inset of Fig.~3(b) of the Letter supports this distinction. ParaToric simulations at $L=12$, $T=0.16$, and $\Delta s=0.01$, initialized from the deconfined phase at small $s$ and the confined phase at large $s$, convert in a single step without a central shelf. Dragged loops probe branch metastability across the coexistence range located in Sec.~\ref{supp:sm:arm-qmc}. Applying the same worldline code to the dual side reproduces the ParaToric branch values within $0.001$--$0.02$ and again gives single-step loops. The wedge is therefore present only on the cluster side.

The frozen manifold is invariant under the one-form symmetry. A plaquette generator generally permutes an individual bare plaquette-$Z$ configuration, but an appropriate mixture over the manifold is weakly symmetric. The exact identity within the all-plus $P_+$ sector does not determine this thermal structure.

\subsection{Numerical determination of Fig.~3(b) of the Letter}
\label{supp:sm:arm-qmc}

The finite-temperature phase boundary of the gauge theory shown in Fig.~3(b) of the Letter is determined with the continuous-time algorithm, $Z$ basis, and validation described in Sec.~\ref{supp:sm:tstar-qmc}. Temperatures and fields are converted to the convention of the Letter by $T=(1-s)\tilde T=\tilde T/(1+h^x)$ and $s=h^x/(1+h^x)$.

\subsubsection{The second-order arm}
At fixed transverse field $h^x=s/(1-s)$ the temperature of the unit-coupling model is scanned across the transition, and $\tilde T_c$ is located by the peak of the specific heat per link computed \emph{directly} as the energy fluctuation with its continuous-time kink-number correction,
\begin{align}
c_V\coloneqq\frac{C}{N_l}
&=\frac{\beta^2\!\left(\langle E^2\rangle-\langle E\rangle^2\right)-n_{\rm kink}}{N_l},
\qquad N_l=3L^3,\label{supp:eq:fluctC}\\
n_{\rm kink}&\coloneqq-\beta\langle H_{\rm off}\rangle .\nonumber
\end{align}
Here $H_{\rm off}\coloneqq-h^x\sum_lX_l$ is the off-diagonal transverse-field term, and $n_{\rm kink}$ is the mean number of field vertices in the worldline configuration. The subtraction in Eq.~(\ref{supp:eq:fluctC}) is required by the continuous-time fluctuation estimator. At $h^x=0$ there are no off-diagonal vertices, and the expression reduces to $c_V=\beta^2\mathrm{Var}(E)/N_l$.

We determine each peak from the mean of eight independent Markov chains using a local three-point parabola, and then extrapolate with the three-dimensional Ising form $\tilde T_{\rm pk}(L)=\tilde T_c+bL^{-1/\nu}$, with $\nu=0.630$~\cite{KPSDV}. Nonparametric resampling of the eight chains at each $(s,L,\tilde T)$ propagates chain-to-chain noise through both the peak interpolation and the finite-size extrapolation. The central fit uses $L=8,10,12,16$. The quoted $1\sigma$ uncertainty combines the bootstrap standard deviation with the maximum shift under $L_{\min}=6,8,10$ and changes of the local interpolation window. The physical temperature is $T_c(s)=(1-s)\tilde T_c$, under which $c_V$ is invariant.

Each chain uses $4\times10^3$ thermalization sweeps followed by $1.6\times10^4$ samples separated by one sweep, where a sweep comprises $3L^3$ single-link updates. Integrated autocorrelation times remain below about $10$ sample units and are monitored throughout. At $h^x=0$, this procedure gives $\tilde T_c(0)=1.323(17)$, consistent with the exact $1.3133$ in Eq.~(\ref{supp:eq:Tc}). We use the exact value only at the $s=0$ endpoint and do not rescale the interior estimates:
\begingroup\footnotesize\setlength{\arraycolsep}{1pt}
\begin{equation}\label{supp:eq:arm-table}
\begin{array}{c|cccccc}
s & 0 & 0.10 & 0.15 & 0.20 & 0.25 & 0.30\\\hline
T_c(s) & 1.3133 & 1.185(3) & 1.113(5) & 1.042(9) & 0.965(8) & 0.887(3)\\[4pt]
s & 0.35 & 0.40 & 0.45 & \mathbf{0.50} & 0.52 & \\\hline
T_c(s) & 0.806(4) & 0.712(12) & 0.607(5) & \mathbf{0.480(6)} & 0.418(6) &
\end{array}
\end{equation}\endgroup
The small-field data provide a direct check of the evenness established in Sec.~\ref{supp:sm:3d-exact}: $\tilde T_c$ decreases by only $1.2\%$ at $h^x=0.25$, with no resolved linear term. A fit to $\tilde T_c(h^x)=\tilde T_c(0)[1-a(h^x)^2]$ over $h^x\le0.6$ determines the leading curvature,
\begin{equation}\label{supp:eq:a-measured}
a=0.21(2),
\end{equation}
with higher-order terms becoming visible only for $h^x\gtrsim0.7$. The measured boundary continues through the self-dual line. Because no symmetry pins a transition to $s=\tfrac12$, the arm crosses it as an ordinary continuous transition at $T_c(\tfrac12)=0.480(6)$ and extends to $s=0.52$.

\subsubsection{The order of the transition along the boundary}
We diagnose the order of the boundary from the volume scaling of the specific-heat maximum in Eq.~(\ref{supp:eq:fluctC}). At a first-order transition, the intensive peak grows as the volume, $c_{V,\rm max}\propto N_l$, and therefore
\(c_{V,\rm max}(L=16)/c_{V,\rm max}(L=8)=8\).
At a three-dimensional Ising transition it grows only as $L^{\alpha/\nu}$, with $\alpha/\nu\simeq0.17$, giving a ratio near $1.13$. For every field from $s=0.10$ to $0.52$, the measured ratios lie between $0.74$ and $1.21$, with mean approximately $1.0$. Statistical noise in the peak height accounts for the scatter around the Ising value, while all ratios remain an order of magnitude below the first-order prediction.

This comparison excludes a strong volume-scaling first-order peak through $s=0.52$, but it neither establishes Ising universality nor rules out a weakly first-order transition by itself. Additional diagnostics in Sec.~\ref{supp:sm:first-order-qmc} support continuity: the slope of the $\langle\tilde{\mathcal M}\rangle(h^x)$ crossing does not sharpen with volume, the susceptibility and specific-heat peaks grow only weakly with $L$---$\chi_{\tilde{\mathcal M}}\approx18.7,33.2,34.1,39.5$ for $L=6,8,10,12$, far below the eightfold volume scaling of a first-order peak---and the Fredenhagen--Marcu ratio~\cite{FredenhagenMarcu} (measured in a separate equilibrium run) shows no jump. The latter is the gauge-invariant ratio of an open half-string amplitude to the square root of the corresponding closed Wilson loop; it jumps at a first-order transition and varies continuously through a second-order one. These tests are consistent with a continuous transition, of the expected three-dimensional Ising type, at the accessible sizes. Binder-cumulant crossings of the one-form order parameter and a data collapse would provide a stronger universality test, but the Letter's ensemble-inequivalence claim depends only on the location of the boundary.

\subsubsection{The first-order segment and the tricritical point}
\label{supp:sm:first-order-qmc}
\begin{figure}[h]
\centering
\includegraphics[width=\columnwidth]{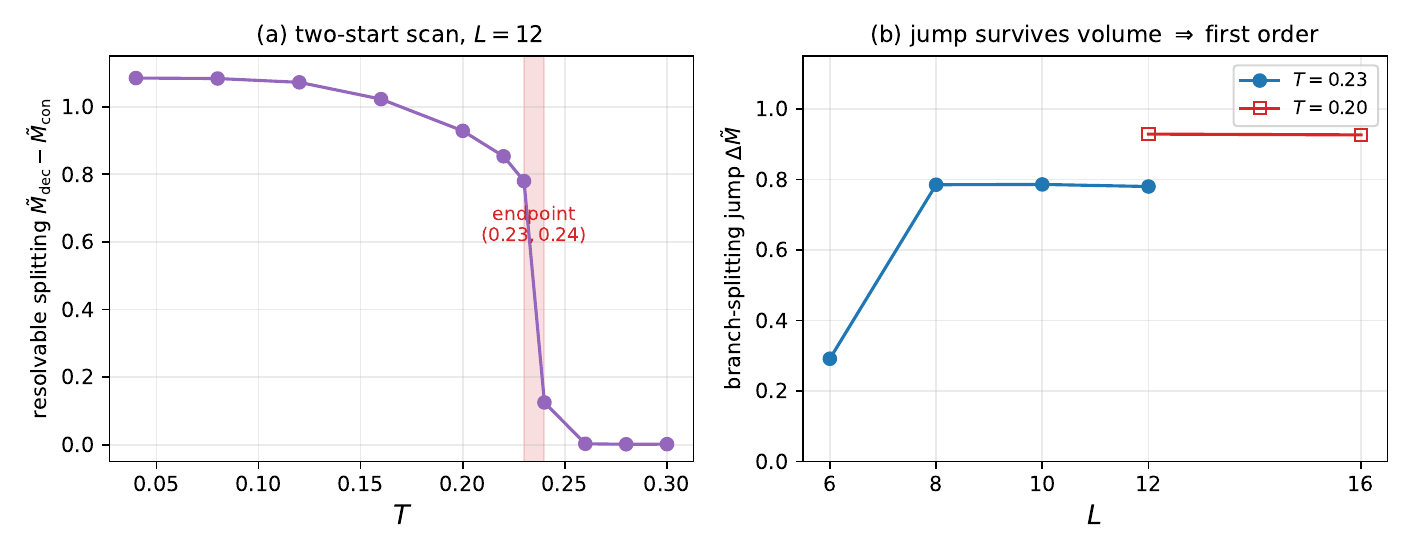}
\caption{Finite-size metastability endpoint of the dual first-order segment, from long-thermalization two-start scans. (a) The resolvable branch splitting $\tilde{\mathcal M}_{\rm dec}-\tilde{\mathcal M}_{\rm con}$ ($L=12$) stays large down to $T=0.23$ and is no longer resolved by $T=0.24$, bracketing the endpoint in $(0.23,0.24)$ (shaded). (b) The splitting \emph{jump} $\Delta\tilde{\mathcal M}(L)$ is volume-independent for $L\ge8$ at $T=0.23$ and at $T=0.20$ ($L=12,16$)---the first-order signature that survives coarse-graining. A crossing \emph{slope} cannot serve this role, because the grid $\Delta s=0.01$ exceeds the finite-size rounding width $\sim T/(N_l\Delta\tilde{\mathcal M})$ and returns a grid-limited, $L$-independent value for either order. This is a finite-size/finite-time bracket, not a thermodynamic tricritical determination.}
\label{supp:fig:tricritical}
\end{figure}

We probe the first-order segment with ParaToric two-branch scans. At each temperature $T$ and field $s$ (grid $\Delta s=0.01$, $L=12$, four seeds) we run two independent simulations that ramp from a deconfined start ($h_{\rm therm}=0.6$) and a confined start ($h_{\rm therm}=1.5$) to the target field $h\coloneqq s/(1-s)$ and then equilibrate---a $60\,000$-sweep prethermalization over a ten-stage ramp followed by $12\,000$ samples at $6$-sweep spacing, about $12\times$ the initial scan. Where the two starts hold opposite branches, $\tilde{\mathcal M}_{\rm dec}>0.12$ and $\tilde{\mathcal M}_{\rm con}<-0.12$ (a branch splitting $\Delta\tilde{\mathcal M}\gtrsim0.24$), we record a resolvable metastable coexistence; its midpoint $s_c(T)$ and half-width $\delta s_{\rm meta}$ form a \emph{metastability} interval bounded by the two spinodals---not an equilibrium wall with a statistical error, since no $s\leftrightarrow1-s$ symmetry pins the equilibrium crossing to the interval center. The long thermalization suppresses but does not remove residual relaxation: near the high-$T$ end of the segment individual seeds still convert within the run. Over $0.04\le T\le0.22$ the strip midpoints, half-widths, and strip-averaged splittings $\Delta\tilde{\mathcal M}$ are
\begingroup\footnotesize\setlength{\arraycolsep}{2pt}
\begin{equation}\label{supp:eq:wall-table}
\begin{array}{c|cccccc}
T & 0.04 & 0.06 & 0.08 & 0.10 & 0.12 & 0.14\\\hline
s_c & 0.510 & 0.510 & 0.515 & 0.515 & 0.525 & 0.530\\
\delta s_{\rm meta} & 0.020 & 0.020 & 0.015 & 0.015 & 0.015 & 0.010\\
\Delta\tilde{\mathcal M} & 1.06 & 1.06 & 1.07 & 1.05 & 1.05 & 1.03\\[4pt]
T & 0.16 & 0.18 & 0.20 & 0.22 & & \\\hline
s_c & 0.535 & 0.535 & 0.540 & 0.540 & & \\
\delta s_{\rm meta} & 0.005 & 0.005 & {<}0.005 & {<}0.005 & & \\
\Delta\tilde{\mathcal M} & 1.01 & 0.97 & 0.93 & 0.85 & &
\end{array}
\end{equation}\endgroup
At $(s,T)=(0.540,0.20)$ the strip-averaged splitting $\Delta\tilde{\mathcal M}=0.93$ at $L=12$ is reproduced at $L=16$ ($0.926$), a $2.4\times$ larger volume [Fig.~\ref{supp:fig:tricritical}(b)]: the \emph{jump} does not shrink with volume, the first-order signature that survives coarse-graining. We stress that the volume test must use this jump and not the crossing \emph{slope}: at $\Delta s=0.01$ the measured $|d\tilde{\mathcal M}/ds|\approx\Delta\tilde{\mathcal M}/\Delta s\approx80$ is grid-limited and $L$-independent for either order, because the finite-size rounding width $\delta s_L\sim T/(N_l\Delta\tilde{\mathcal M})\sim10^{-4}$ ($N_l=3L^3$) lies far below the grid. The branch plateaus are independent of the preparation field. The identical protocol at $T=0.32$ resolves no metastable coexistence: the two starts converge, with a residual splitting below $0.17$.

For $T\ge0.24$ the two starts converge and the scans resolve no stable two-branch splitting at the accessible run length; we then read the boundary \emph{location} from the equilibrium crossing $\langle\tilde{\mathcal M}\rangle(s)=0$, which sits at $s_c\simeq0.544$ for $T=0.24$--$0.32$. Here $\langle\tilde{\mathcal M}\rangle=\partial_s f=0$ need not mark a transition---it is where the free-energy slope vanishes and can fall inside a single phase---so it estimates the wall position but does not by itself certify the order there. The continuous character in this range is inferred from the specific-heat arm of the preceding subsection, whose boundary connects to this locus, rather than from these crossings.

The resolvable metastable coexistence, carrying a large volume-independent jump [Fig.~\ref{supp:fig:tricritical}], therefore persists down to $T=0.23$ ($\Delta\tilde{\mathcal M}=0.93,0.85,0.78$ at $T=0.20,0.22,0.23$) and is no longer resolved by $T=0.24$, where the maximum splitting has fallen to $0.12$ and is itself partly a within-run transient. This brackets the finite-size metastability endpoint of the segment,
\begin{equation}\label{supp:eq:tricritical-sm}
s^*\simeq0.542,\qquad 0.23<T^*_{\rm FS}<0.24,
\end{equation}
a finite-size, finite-time bound rather than a thermodynamic tricritical point: on this weakly first-order segment the metastability barrier is too small to separate metastable lifetimes from critical slowing down at the accessible sizes (\S\ref{supp:sm:first-order-qmc}, methodological remarks below), and pinning the thermodynamic $T^*$ would require multicanonical or replica-exchange sampling together with a phase-sensitive diagnostic (Fredenhagen--Marcu, Binder, or susceptibility). The segment connects continuously to the specific-heat arm above the bracket and bends toward the exact endpoint $(\tfrac12,0)$ below it.
Over $0.04\le T\le0.22$ the strip midpoint moves from $s_c\simeq0.540$ near the endpoint to $\simeq0.510$ at the lowest simulated temperature $T=0.04$, consistent within the metastability interval with the exact zero-temperature endpoint $s=\tfrac12$ anticipated in Sec.~\ref{supp:sm:3d-wall} [$\lim_{T\to0}s_c(T)=\tfrac12$]; by $T=0.24$ no strip is resolved. The metastability half-width widens only mildly, from below $0.005$ near the endpoint to $0.020$ at $T=0.04$; because it is a dynamical interval and its midpoint is not the equilibrium wall, the midpoints are not used to extract a charge gap.

A fixed-$s$ temperature scan at $s=0.540$, directly over the segment, explains why the peak method of the preceding subsection fails there. The scan crosses the coexistence line at a grazing angle, so $d\langle E\rangle/d\tilde T$ has only a broad, smeared bump with no volume scaling. Meanwhile the integrated autocorrelation time rises to $\tau\simeq153$ at $L=12$, compared with $\lesssim7$ away from the line. The resulting slow two-phase dynamics serves as the wall diagnostic.

\subsubsection{Methodological remarks}
Metastability alone would misplace the tricritical point. The barrier on this weakly first-order segment is too small to separate metastable lifetimes cleanly from critical slowing down at the accessible sizes. Scalar histogram summaries are also unreliable: an equal-weight criterion for $P(\tilde{\mathcal M}>0)$ or peak separations from conditional medians can mistake a broad single-phase distribution centered near $\tilde{\mathcal M}=0$ for coexistence, because $\tilde{\mathcal M}$ is not a symmetry order parameter here.

We therefore use the volume scaling of $c_{V,\rm max}$ for the continuous arm. For the first-order segment, we require a volume-independent branch-splitting jump, preparation independence, and the disappearance of the resolvable splitting at both ends. Figure~3(b) of the Letter combines the arm in Eq.~(\ref{supp:eq:arm-table}), the segment in Eq.~(\ref{supp:eq:wall-table}), the finite-size metastability endpoint Eq.~(\ref{supp:eq:tricritical-sm}) (star), the exact zero-temperature endpoint $(\tfrac12,0)$~\cite{ReissSchmidt,ParaToric}, and the tangent $T_c(0)(1-s)$ from Sec.~\ref{supp:sm:3d-exact}. In the interval $0.24\lesssim T\lesssim0.38$, the grazing crossing invalidates the peak method and the scans resolve no metastable coexistence; there we locate the boundary from the equilibrium $\langle\tilde{\mathcal M}\rangle(h^x)$ crossings, whose order is not determined by these crossings but inherited from the continuous arm they join.

\end{document}